\documentclass[useAMS,usenatbib]{mn2e}
\pdfoutput=1
\usepackage[pdftex]{graphicx}
\usepackage{natbib}
\usepackage{hyperref} 
\hypersetup{pdfpagemode=UseNone} 

\newcommand{\msun}{M$_{\odot}\,$}

\newcommand{\apj}{ApJ}
\newcommand{\apjl}{ApJL}
\newcommand{\apjs}{ApJS}
\newcommand{\aap}{A\&A}
\newcommand{\aapr}{A\&ARv}
\newcommand{\araa}{ARA\&A}
\newcommand{\mnras}{MNRAS}
\newcommand{\physrep}{Physics Reports}
\newcommand{\nat}{Nature}

\title[The structure of halo gas]{Zooming in on accretion - I. The structure of halo gas}
\author[D. R. Nelson et al.]{Dylan Nelson$^{1}$\thanks{E-mail: dnelson@cfa.harvard.edu},
Shy Genel$^{1,2}$\thanks{Hubble Fellow.}, Annalisa Pillepich$^{1}$, Mark Vogelsberger$^{3}$, \newauthor Volker Springel$^{4,5}$, Lars Hernquist$^{1}$\\\\
$^{1}$Harvard-Smithsonian Center for Astrophysics, 60 Garden Street, Cambridge, MA, 02138, USA\\
$^{2}$Department of Astronomy, Columbia University, 550 West 120th Street, New York, NY, 10027, USA\\
$^{3}$Kavli Institute for Astrophysics and Space Research, Department of Physics, MIT, Cambridge, MA, 02139, USA\\
$^{4}$Heidelberg Institute for Theoretical Studies, Schloss-Wolfsbrunnenweg 35, 69118 Heidelberg, Germany\\
$^{5}$Zentrum f\"{u}r Astronomie der Universit\"{a}t Heidelberg, ARI, M\"{o}nchhofstr. 12-14, 69120 Heidelberg, Germany\\
}

\begin{document}

\maketitle

\begin{abstract}
We study the properties of gas in and around $10^{12}$\,\msun haloes at $z\!=\!2$ using a suite of high-resolution 
cosmological hydrodynamic `zoom' simulations. We quantify the thermal and dynamical structure of these gaseous 
reservoirs in terms of their mean radial distributions and angular variability along different sightlines. 
With each halo simulated at three levels of increasing resolution, the highest reaching a baryon mass resolution of 
$\sim$10,000 solar masses, we study the interaction of filamentary inflow and the quasi-static hot halo atmosphere.  
We highlight the discrepancy between the spatial resolution available in the halo gas as opposed to within the galaxy 
itself, and find that stream morphologies become increasingly complex at higher resolution, with large coherent flows 
revealing density and temperature structure at progressively smaller scales. 
Moreover, multiple gas components co-exist at the same radius within the halo, making radially averaged analyses 
misleading. This is particularly true where the hot, quasi-static, high entropy halo atmosphere interacts with cold, 
rapidly inflowing, low entropy accretion. We investigate the process of gas virialization and identify different 
regimes for the heating of gas as it accretes from the intergalactic medium. Haloes at this mass have a 
well-defined virial shock, associated with a sharp jump in temperature and entropy at $\ga$\,1.25\,$r_{\rm vir}$. 
The presence, radius, and radial width of this boundary feature, however, vary not only from halo to halo, but also 
as a function of angular direction, covering roughly $\sim$\,85\% of the 4$\pi$ sphere. 
Our findings are relevant for the proper interpretation of observations pertaining to the circumgalactic medium, 
including evidence for large amounts of cold gas surrounding massive haloes at intermediate redshifts.
\end{abstract}



\section{Introduction}

Initially following the gravitational collapse of a dark matter overdensity, gaseous haloes subsequently grow through 
the accretion of baryons from the intergalactic medium (IGM). Their evolving structure across cosmic time has been the 
subject of theoretical as well as observational interest for several decades. As the transitional state between the 
diffuse IGM and the star-forming interstellar medium of galaxies, these gas reservoirs regulate the stellar growth of 
forming galaxies. Understanding not only the structure of halo gas, but also its origin and subsequent evolution, is 
therefore essential for any comprehensive theory of galaxy formation.


Although the accretion of gas will follow that of dark matter, the presence of additional physical processes including 
hydrodynamical forces and radiative cooling imply additional complexity for the acquisition of baryons. In the classic 
picture, gas accreting from the intergalactic medium will shock heat to the virial temperature of the halo. If the 
radiative cooling timescale is sufficiently long, it will then form a hot, pressure supported atmosphere in approximate 
equilibrium \citep{rees77,silk77,wr78}. Cooling can proceed, delivering gas into the halo centre \citep{wf91}. For 
sufficiently low mass haloes, this timescale will be short enough that a `stable virial shock' cannot develop, and gas 
accretion from the IGM will proceed as rapidly as dynamically allowed \citep{bd03}. However, this theoretical foundation 
is largely based on a one-dimensional picture, while dark matter haloes and their gaseous counterparts are decidedly not 
spherically symmetric. Numerical hydrodynamical simulations have indicated that galaxies can acquire their gas in a 
fundamentally different manner \citep{katz03,abadi03,keres05,ocvirk08}. In particular, finding that coherent, filamentary 
inflows can fuel star formation in a central galaxy while avoiding shock heating to the virial temperature. Such streams 
arise naturally from the topology of large-scale structure, particularly at high redshifts of $z \ga 2$ 
\citep{dekel09,agertz09,keres09,danovich12}.

Gas inflow has been studied in the context of many key questions in galaxy formation, most fundamental of which is 
perhaps its link to star formation \citep{opp10,gabor14,almeida14} and the process of quenching 
\citep{birnboim03,birnboim07,gabor12,feldmann15,aragon15}. The accretion of cosmological gas will leave a fundamental 
imprint on the thermal and dynamical properties of the quasi-static halo gas, as will outflows from energetic feedback 
processes in galaxies \citep{putman12}. By studying the properties of this halo gas we can investigate the interplay of 
inflows and outflows with metals \citep[e.g.][]{shen12,hummels13,ford14} as well as with neutral hydrogen 
\citep[e.g.][]{fgk11,fg14,fum14}.


The state of gas in and around galaxy haloes has also received significant observational scrutiny in the local universe 
as well as at $z\!\sim\!2$, near the peak of the cosmic star formation rate. At these high redshifts both hydrogen and 
metals are accessible as absorption signatures in sightlines towards background objects, enabling a probe of the 
interaction between galaxies and the IGM in the vicinity at this period of rapid stellar growth. Efforts include the 
`quasars probing quasars' series \citep{hennawi06,prochaska13,prochaska14b}, the Keck Baryonic Structure Survey 
\citep{steidel10,rudie12,rudie13,turner14a}, and other 2\,$\la$\,$z$\,$\la$\,3 studies 
\citep{simcoe06,pieri14,rubin14,crighton15}. Collectively they probe the covering fractions, radial profiles, and 
kinematics of {\small HI} and many metal ions including those of oxygen, carbon, neon, silicon, and magnesium.

Perhaps the most puzzling discovery which remains at the present unreconciled with theory is the presence of a large 
amount of cold ($\sim$\,10$^4$\,K), metal-enriched gas widely distributed at least out to the virial radius of massive 
(10$^{12}$\,$\sim$\,10$^{12.5}$\,\msun) haloes. For instance, \cite{prochaska14b} find a covering fraction approaching 
unity for strong {\small C II} absorption at a hundred physical kiloparsecs or more from the halo centre. The origin of 
this gas and the process by which it is either maintained or replenished within the hot halo remains uncertain. 
Consequently, this regime provides a powerful test-bed for current hydrodynamic simulations of galaxy formation. Not only 
in terms of the physical modelling of feedback and the resultant galactic-scale outflows \citep{muratov15}, but more 
fundamentally in terms of our ability to resolve the gas-dynamical processes and the spatial scales relevant for the 
physics of cosmological gas accretion.


This paper investigates the thermal and dynamical structure of halo gas in eight simulated $\simeq$\,10$^{12}$\,\msun 
haloes at $z\!=\!2$. In Section \ref{sMethods} we describe the simulation technique and analysis methodology. Section 
\ref{sResVis} addresses the issue of resolution for halo gas and presents a visual overview of the systems. Section 
\ref{sPhys} considers the radially averaged gas properties, while Section \ref{sAngVar} expands this analysis to explore 
the angular variability of halo structure. We discuss our results in the context of observations of the gas 
content of haloes in Section \ref{sDiscussion} and summarize our conclusions in Section \ref{sConclusions}.


\section{Methods} \label{sMethods}

\subsection{Initial Conditions} \label{ssICs}

\begin{table*}
  \caption{General characteristics of our three resolution levels, L9, L10, and L11. First, the effective resolution of 
  an equivalent uniform box. Next, the mean number of high resolution gas elements, number of timesteps, baryonic mass 
  resolution, and dark   matter mass resolution. The Plummer equivalent comoving gravitational softening lengths, and 
  their physical values at $z\!=\!2$. The minimum gas cell spatial size in physical parsecs at $z\!=\!2$, and the mean 
  gas cell spatial size in the halo, between 0.5\,$r_{\rm vir}$ and 1.0\,$r_{\rm vir}$, in physical kiloparsecs at $z\!=\!2$.}
  \label{tResLevels}
  \begin{center}
    \begin{tabular}{llllllllll}
     \hline 
 Res & 
 N$_{\rm part}^{\rm eff}$ & 
 N$_{\rm part}^{\rm HR}$ & 
 $\Delta$t [\#]&
 m$_{\rm baryon}$ [\msun] & 
 m$_{\rm DM}$ [\msun] & 
 $\epsilon_{\rm grav}^{\rm comoving}$ [pc] &
 $\epsilon_{\rm grav}^{\rm z\!=\!2}$ [pc] &
 $r_{\rm cell}^{\rm min}$ [pc] &
 $r_{\rm cell}^{\rm halo}$ [kpc] \\ \hline\hline
        
 L9  & $512^3$  & 800,000    & 80,000  & 1.0 x 10$^6$  & 5.1 x 10$^6$  & 1430 & 480 & 31  & 2.7 \\
 L10 & $1024^3$ & 7,000,000  & 260,000 & 1.3 x 10$^5$  & 6.4 x 10$^5$  & 715  & 240 & 11  & 1.6 \\
 L11 & $2048^3$ & 64,000,000 & 870,000 & 1.6 x 10$^4$  & 8.0 x 10$^4$  & 357  & 120 & 3.3 & 0.8 \\ \hline
    \end{tabular}
  \end{center}
\end{table*}

All simulations evolve initial conditions which are a random realization of a WMAP-9 consistent cosmology 
($\Omega_{\Lambda,0}=0.736$, $\Omega_{m,0}=0.264$, $\Omega_{b,0}=0.0441$, $\sigma_8=0.805$, $n_s=0.967$ and $h=0.712$). 
We use the {\small MUSIC} code \citep[][v1.5, r375]{hahn11} to generate multi-mass `zoom' ICs, under the 2LPT approximation and with a tabulated 
transfer function from {\small CAMB} \citep{lewis00}. First, we evolve a low resolution, dark matter only uniform 
periodic box of side-length 20$h^{-1}$ Mpc $\simeq 28.6$ Mpc with $128^3$ particles (`L7', where L$N$=2$^N$), from a 
starting redshift of $z\!=\!99$ down to $z\!=\!2$. At this redshift, there are 20 haloes with total mass between 
$10^{11.8}$\msun\, and $10^{12.4}$\msun\, from which we choose eight at random to re-simulate at higher resolution. We 
do not select for any additional properties -- e.g., merger history or environment. All particles within some factor 
of the virial radius of each selected halo (ranging from 3.6\,$r_{\rm vir}$ to 7.0\,$r_{\rm vir}$ in all cases, see 
\cite{onorbe14} for relevant considerations) are identified at $z\!=\!2$. We take $r_{\rm vir}$\,=\,$r_{\rm 200,crit}$ 
the radius enclosing a mean overdensity 200 times the critical density. This factor was chosen by trial and error with 
evaluation of contamination levels in low resolution test runs. The convex hull of the $z\!=\!99$ positions of all selected 
particles is then used to define the high resolution refinement region.

For each halo, new initial conditions are generated for each of L9, L10, and L11, corresponding to $512^3$, $1024^3$, 
and $2048^3$ total particles if the parent box were to be simulated at a uniform resolution. We note that the mass 
resolution of L11 is between Aquarius levels `3' and `4' \citep[of e.g.][]{marinacci14a,scan12a}, while the resolution 
of L9 is approximately equal to the resolution in modern, large volume cosmological simulations. Baryons 
are included by splitting each dark matter particle according to the cosmological baryon fraction, into one DM particle 
and one gas cell, such that the centre of mass position and velocity are preserved. Therefore we do not consider a 
separate transfer function for the baryonic component. The files required to generate our initial conditions, including 
the {\small CAMB} transfer function, noise seeds, and convex hull point sets, are made publicly available 
online\footnote{\url{http://www.illustris-project.org/files/Nelson15b_ICs.zip}}. The fundamental characteristics of each 
resolution level are given in Table \ref{tResLevels}, while the physical properties and numerical details for each of the 
eight haloes are detailed in Table \ref{tHaloes}. 

\begin{table}
  \setlength{\tabcolsep}{3pt}
  \caption{Details on the eight simulated haloes: the total halo mass and (physical) virial radius at $z\!=\!2$ from the 
  parent box. The radius of the enclosing sphere at $z\!=\!2$ used to define the Lagrangian region, in terms of 
  $r_{\rm vir}$, and the number of high resolution elements, for each of dark matter and gas. Both are listed for the 
  L11 level only. The minimum radius from the halo centre reached by (i) contaminating low resolution dark matter 
  particles and (ii) Monte Carlo tracers originating in low resolution gas cells, in units of $r_{\rm vir}$. The total 
  number of timesteps to reach $z\!=\!2$.}
  \label{tHaloes}
  \begin{center}
    \begin{tabular}{ccrcrccc}
     \hline 
 Halo &
 $M_{\rm halo}^{\rm par}$ &
 $r_{\rm vir}^{\rm par}$ &
 $r_{\rm HR}^{\rm L11}$ &
 $N_{\rm HR}^{\rm L11}$ &
 $r_{\rm LR}^{\rm min}$ & 
 $r_{\rm LR}^{\rm min}$ & 
 $\Delta t$ \\
 \# &
 [log \msun] &
 [kpc] &
 [$r_{\rm vir}$] &
 [10$^6$] &
 [$r_{\rm vir}$]$_{\rm dm}$ &
 [$r_{\rm vir}$]$_{\rm tr}$ & 
 \# \\ \hline\hline
        
 h0 & 12.1 & 114 & 3.6 & 70.0  & 1.77 & 2.16 & 829714 \\ 
 h1 & 12.1 & 104 & 4.8 & 66.7  & 2.12 & 2.75 & 701681 \\ 
 h2 & 11.9 & 92  & 6.0 & 24.2  & 2.83 & 3.02 & 955189 \\ 
 h3 & 11.9 & 96  & 7.0 & 33.9  & 2.74 & 3.23 & 812983 \\ 
 \hline
 h4 & 12.0 & 103 & 6.0 & 68.4  & 2.13 & 2.89 & 861224 \\ 
 h5 & 12.0 & 103 & 4.2 & 59.9  & 1.04 & 1.04 & 931088 \\ 
 h6 & 12.1 & 97  & 4.8 & 74.4  & 1.32 & 1.59 & 980918 \\ 
 h7 & 11.9 & 94  & 4.4 & 52.8  & 0.94 & 1.93 & 866242 \\ 
 \hline
    \end{tabular}
  \end{center}
\end{table}

\subsection{Simulation Code and Physics} \label{ssSimulations}

We employ the {\small AREPO} code \citep[][r25505]{spr10} to solve the coupled equations of ideal continuum hydrodynamics 
and self-gravity. An unstructured, moving, Voronoi tessellation of the domain provides the spatial discretization for 
Godunov's method with a directionally un-split MUSCL-Hancock scheme \citep{vl77} and an exact Riemann solver, obtaining 
second order accuracy in space. Since we allow the mesh generating sites to move, with a velocity equal to the local fluid 
velocity field modulated by corrections required to maintain the regularity of the mesh, this numerical approach would 
be classified as an Arbitrary Lagrangian-Eulerian (ALE) scheme.
Gravitational accelerations are computed using the Tree-PM approach, where long-range forces are calculated with a 
Fourier particle-mesh method, medium-range forces with a hierarchical tree algorithm \citep{bh86}, and short-range 
forces with direct summation \citep[as in][]{spr05b}.
A local, predictor-corrector type, hierarchical time stepping method obtains second order accuracy in time.
Numerical parameters tangential to our current investigation -- for example, related to mesh regularization or 
gravitational force accuracy -- are detailed in \cite{spr10} and \cite{vog12}. 

We include a redshift-dependent, spatially uniform, ionizing UV background field \citep{fg09}. Gas loses internal energy 
from optically thin radiative cooling assuming a primordial H/He ratio \citep{katz96}. The production of metals and 
metal line cooling contributions are not included.Star formation and the associated ISM pressurisation from unresolved 
supernovae are included with an effective equation of state modelling the ISM as a two-phase medium, following \cite{spr03}. 
Gas elements are stochastically converted into star particles when the local gas density exceeds a threshold value of 
$n_{\rm H}=0.13$ cm$^{-3}$. Furthermore, there is no resolved 
stellar feedback that would drive galactic-scale winds, nor any treatment of black holes or their associated feedback. 
All of the simulations considered in this work disregard the possible effects of radiative transfer, magnetic fields, 
and cosmic rays. The set of implemented physics is essentially identical to the `moving mesh cosmology' paper series 
as described in \cite{nelson13}.

We identify dark matter haloes and their gravitationally bound substructures using the {\small SUBFIND} algorithm 
\citep{spr01,dolag09} which is applied on top of a friends-of-friends cluster identification. We refer to the most 
massive substructure in each FoF group as the halo itself. Merger trees are constructed using the {\small SUBLINK} 
code \citep{rodrig15} to link haloes to their progenitors and descendants at different points in time.

All runs include the Monte Carlo tracer particle technique \citep{genel13} in order to follow the evolving properties of 
gas elements over time, with five tracers per initial gas cell. This is a probabilistic method, where tracers are 
exchanged between parents based explicitly on the corresponding mass fluxes. By locating a subset of their unique IDs at 
each snapshot we can, by reference to their parents at that snapshot, reconstruct their spatial trajectory or 
thermodynamic history. 

In the current work we make limited use of the assembly and accretion histories of the haloes, as determined with the tracer 
particles and merger trees, respectively. Specifically, we verify the correspondence between haloes at different 
resolution levels in order to comment on the convergence properties of our simulations. We reserve as future work -- the 
second paper in this series -- a quantitative analysis of the rates and modes of accretion and the impact of merging 
substructures.


\section{Resolution Considerations and Visual Inspection} \label{sResVis}

In the vast majority of galaxy formation simulations, spatial resolution is naturally adaptive and follows the 
hierarchical clustering of structure formation in $\Lambda$CDM. This is true for the dark matter, where the 
Vlassov-Poisson equations are solved with a Monte-Carlo approach. The N-body method is typically used 
\citep[for notable exceptions see][]{yoshikawa13,hahn15}. This natural adaptivity also holds for the gaseous component 
in particle methods like SPH, adaptive grid methods with the typical density-based refinement criteria 
\citep[for notable exceptions see][]{iapichino08,rosdahl12}, and the moving-mesh code used in this work. The resolution 
is therefore better in collapsed structures than it is in under-dense regions, and maximal in galaxies themselves. The 
gaseous haloes surrounding galaxies are poorly resolved in comparison, to the extent that modern cosmological simulations 
\citep[e.g.][]{khandai14,dubois14,vog14a,schaye15} which just barely resolve the internal structure of individual 
galaxies may be insufficiently capturing gas-dynamical processes in the halo.

\begin{figure}
\centerline{\includegraphics[angle=0,width=3.4in]{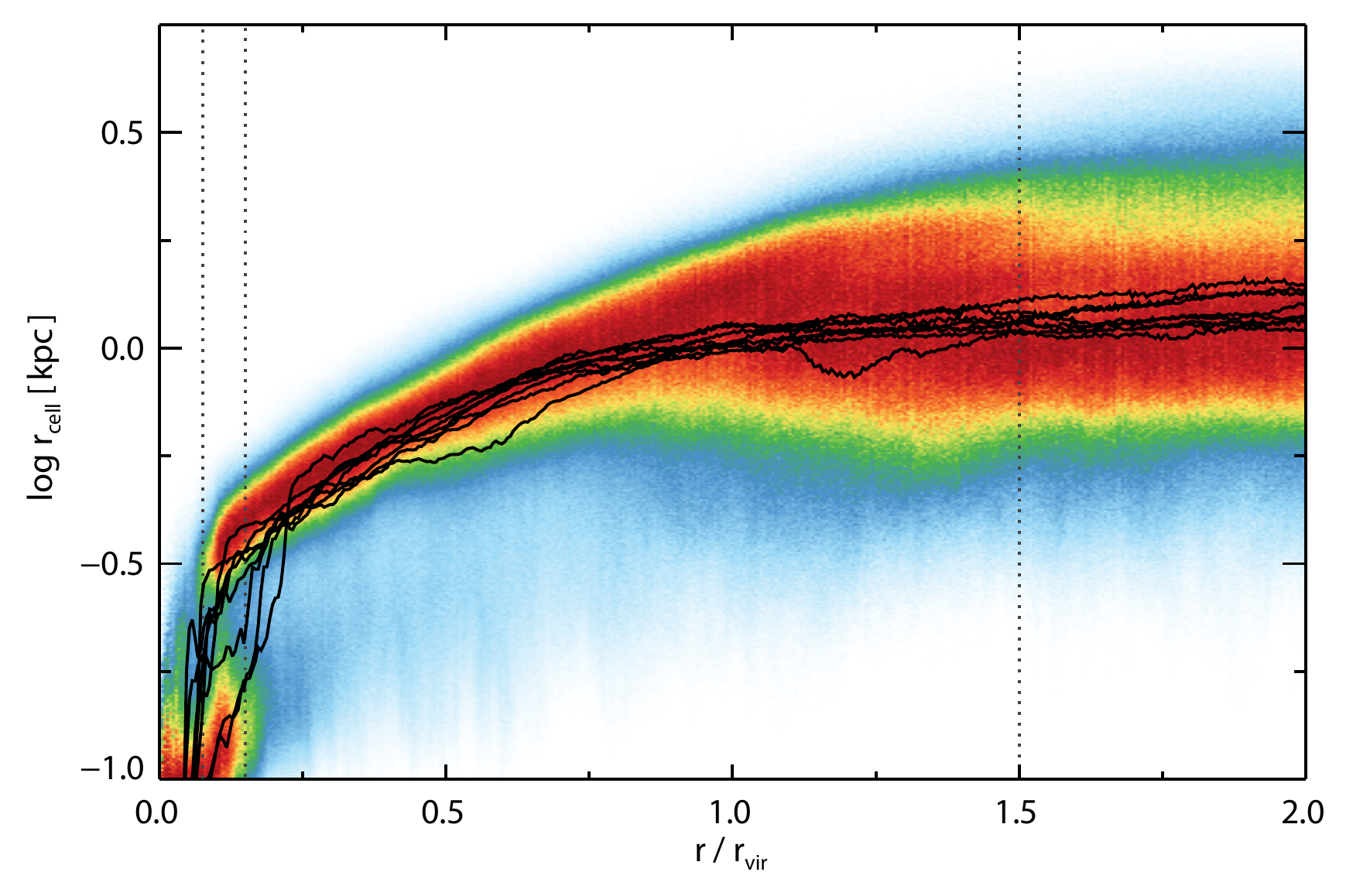}}
\caption{ The mass-weighted distribution of gas spatial resolution as a function of radius, for all eight halos stacked 
together at L11 resolution. As a proxy for spatial resolution, we show the gas cell sphere-equivalent radii $r_{\rm cell}$. Each radial bin 
is normalized independently, such that the colour mapping reaches its maximum intensity at each radius, independent of 
the radial distribution of gas mass. The dark vertical lines indicate $1.5$\,$r_{\rm vir}$, $0.15$\,$r_{\rm vir}$ and 
$0.07$\,$r_{\rm vir}$, the last two bounding the radial range where a density bimodality arises due to the central galaxy. 
Substructures are excised.
 \label{fig_cellsize_vs_rad}} 
\end{figure}

We begin by quantifying the spatial resolution of gas in the halo. Figure \ref{fig_cellsize_vs_rad} shows the 
mass-weighted two dimensional distribution of cell size $r_{\rm cell}$ as a function of radius. As a proxy for the 
irregular shape of Voronoi polyhedra, we take the sphere-equivalent radius $r_{\rm cell} = (3V_{\rm cell} / 4\pi)^{1/3}$ 
where $V_{\rm cell}$ is the exact Voronoi cell volume. The mesh regularization scheme ensures that this value is 
approximately equal to the actual geometrical distance between the generating point and each face centre. All haloes are 
stacked together at the highest (L11) resolution level, while black lines show the median relation for each halo separately.

Beyond $r/r_{\rm vir} > 1.25$ the hydrodynamic resolution is nearly constant, but with a broad distribution spanning from 
$\sim$\,500\,pc to $\sim$\,3\,kpc (physical), depending on what level of overdensity the gas cell resides in. In the halo 
region, $0.25 < r/r_{\rm vir} < 1.0$ the mean resolution scales from $\sim$\,400\,pc up to $\sim$\,1\,kpc. The inner halo, 
typically bounded by $r \la 0.15 r_{\rm vir}$ sees the mean cell size decrease rapidly under $\sim$\,100\,pc as dense 
gas features associated with the central galaxy begin to dominate. We return to the implications of the relatively coarse 
halo resolution in the discussion.

\begin{figure*}
\centerline{\includegraphics[angle=0,width=7.2in]{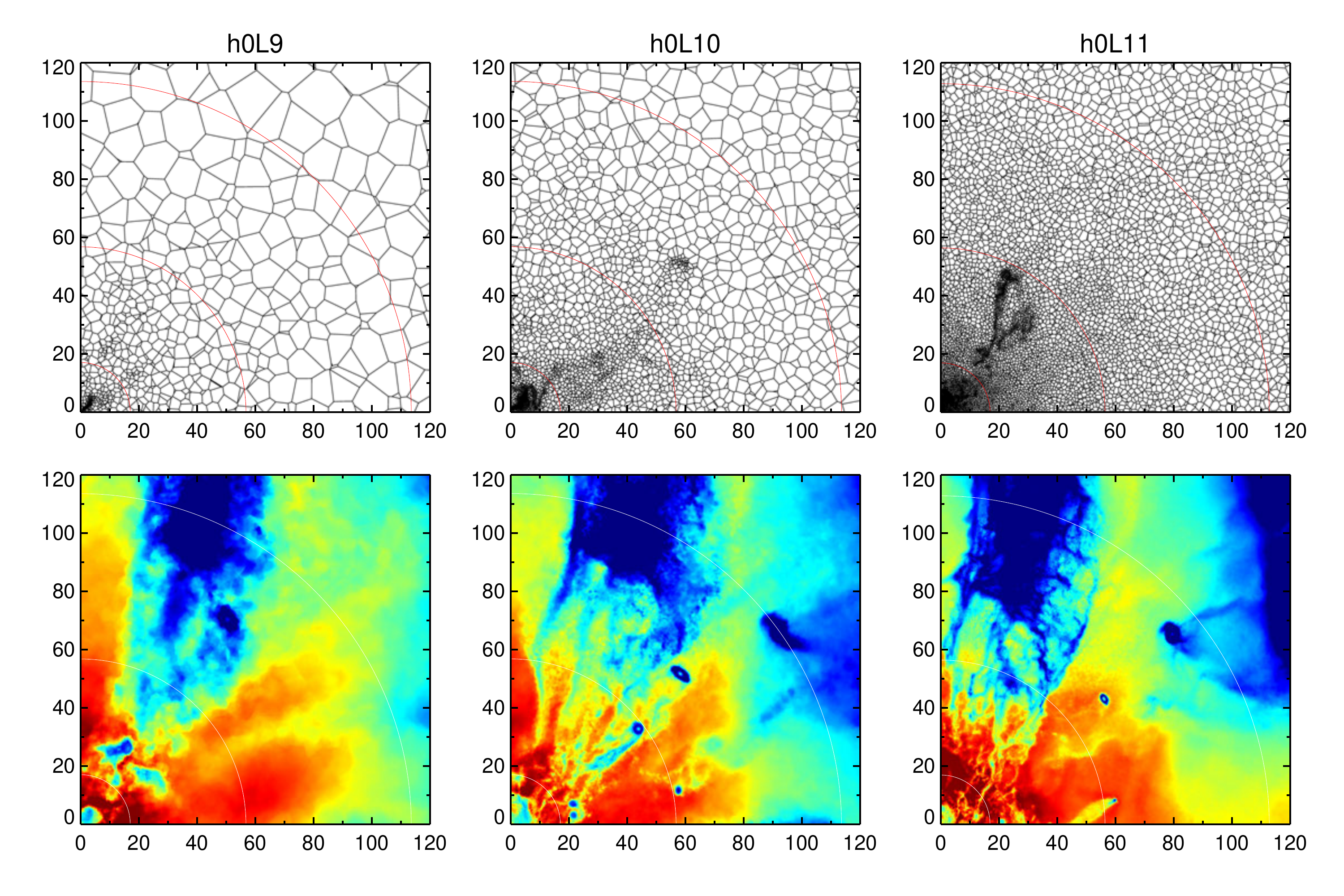}}
\caption{ Slice through the Voronoi mesh showing the underlying spatial discretization (top) and a projection through 
the linearly reconstructed, mass-weighted gas temperature field (bottom) with $\sim$\,$10^4$\,K as blue and 
$\sim$\,$10^6$\,K as red. Halo h0 is shown at $z\!=\!2$. We include only the upper right quadrant, after centring the 
halo at the origin, in orthographic projection. The circles denote \{0.15,0.5,1.0\}\,$r_{\rm vir}$, the axes units are 
in physical kpc, and the projection depth is 238 kpc. The average gas cell size near the virial radius drops with 
increasing spatial resolution (from left to right), from $\sim$2.7 kpc to $\sim$0.8 kpc physical. Note that the exact 
state of gas in the halo, and the location of satellites, differs somewhat due to timing offsets between the three runs.
 \label{fig_mesh_slice_res}} 
\end{figure*}

In Figure \ref{fig_mesh_slice_res} we show a visual example of the resolution issue and a first look at the impact of 
higher hydrodynamical resolution in the halo regime. One halo (h0) is included at the three increasing zoom 
levels considered in this work, L9, L10, and L11, increasing from left to right. On the top row we plot a slice of the 
Voronoi mesh used to evolve the gas component.\footnote{Note that the intersection of a plane with the three-dimensional 
Voronoi tessellation does not in general produce a two-dimensional Voronoi graph. Therefore the shapes of individual 
cells, as shown, do not reflect the actual regularity of the computational mesh.} The slice is centred on the central 
galaxy, which lies at the origin, and to focus on details we include only the upper-right quadrant. The outer red circle 
marks the virial radius of $114$ physical kpc at $z\!=\!2$. At all three resolution levels we see that the high-density 
gas near the origin is resolved at the sub-kpc level and impossible to separate at this scale, while at the virial 
radius individual gas elements are substantially larger. At L9, characteristic of modern cosmological simulations, the 
virial arc across a quadrant is only resolved by $\simeq$\,20 cells, while the radial structure between 
0.5\,$r_{\rm vir}$ and 1.0\,$r_{\rm vir}$ is only resolved by $\simeq$\,10 cells. This is roughly consistent with a 
zeroth order estimate, assuming uniform density, for a number of cells per dimension of

\begin{equation}
N_{\rm cell,dim} \sim \left( \frac{ (\Omega_b / \Omega_m) \times M_{\rm halo}}{m_{\rm cell}} \right)^{1/3} \simeq 50
\end{equation}

\noindent which are available to resolve the halo gas structure of a $10^{12}$\,\msun halo. At L11 this increases to 
$N_{\rm cell,dim} \simeq 250$, but the scaling is obviously slow in three-dimensions. In the discussion we consider a 
possible method for improving upon this resolution issue in the halo with future work.

Across the bottom row we show a projection of mass-weighted gas temperature, where the colormap extends from cold, 
$\sim$\,$10^4$\,K gas as blue to hot, $\sim$\,$10^6$\,K gas as red. Some interesting differences emerge as the gas 
resolution in the halo increases. The large cold inflow which begins to experience substantial heating at 
$(0.4 - 0.5) \times r_{\rm vir}$ shows some temperature inhomogeneities at L9, but is essentially a single coherent 
structure. At L10, this inflow is resolved into a number of smaller gas filaments by the time it crosses half the virial 
radius. At L11 the temperature structure becomes even more complex, throughout the halo, with small scale features 
emerging below the cell-size of the lower resolution runs. Clearly the morphology of accreting gas and its interaction 
with the quasi-static hot halo material depends on numerical resolution to some degree. In the following sections we 
further investigate this dependence by analysing the structure of halo gas at our highest available resolution.

\subsection{Visual Inspection}

\begin{figure*}
\centerline{\includegraphics[angle=0,width=7.0in]{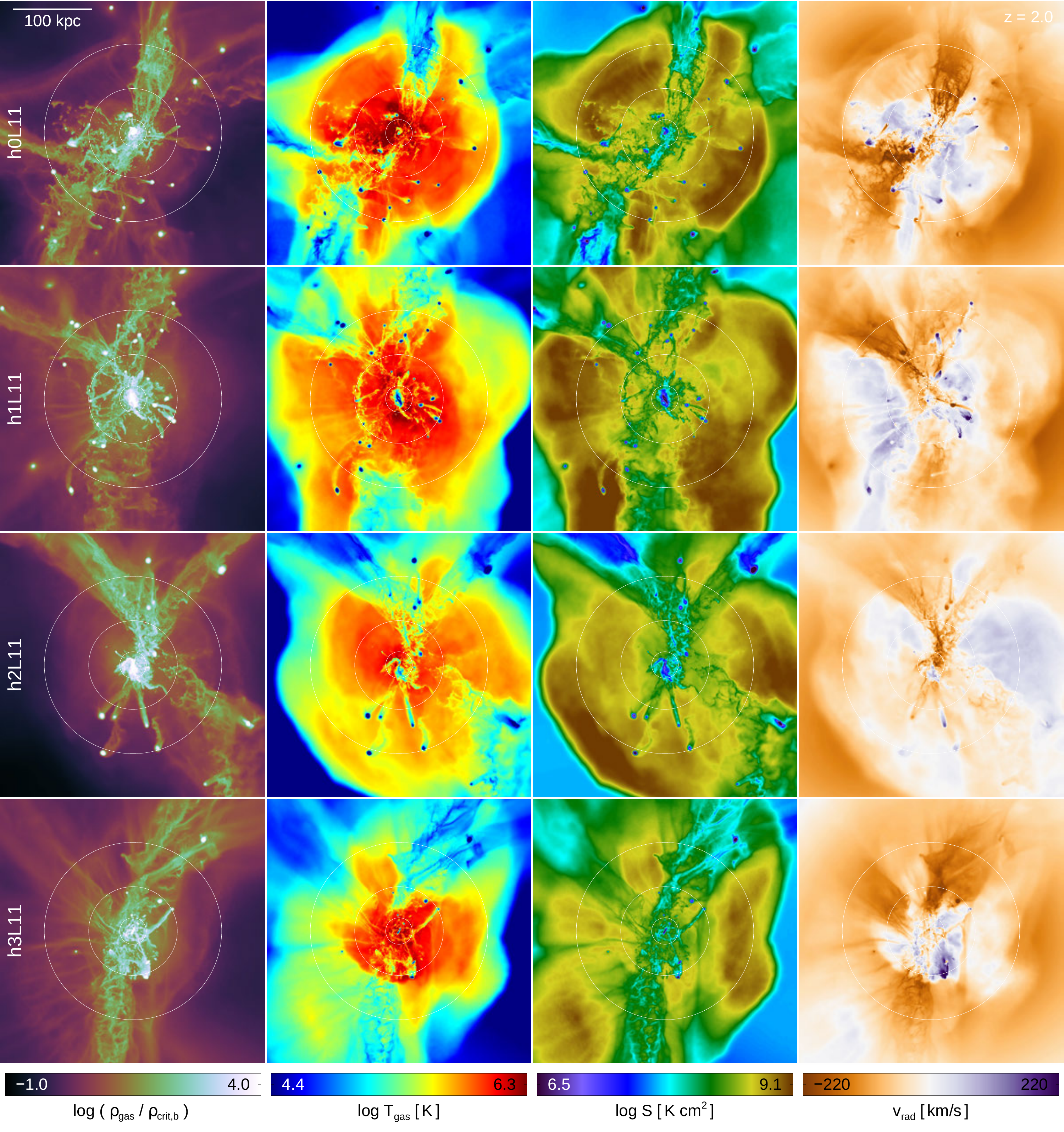}}
\caption{ Mass-weighted projections of gas density, temperature, entropy, and radial velocity for the first four simulated 
haloes at $z\!=\!2$. In each case, all gas cells within a cube of side-length 3\,$r_{\rm vir}$ are included, and 
distributed using the standard cubic spline kernel with $h\,=\,2.5\,r_{\rm cell}$ in orthographic projection. The white 
circles denote \{0.15,0.5,1.0\}\,$r_{\rm vir}$. Gas density is normalized to the critical baryon density at $z\!=\!2$. The 
temperature of star forming gas on the eEOS is set to a constant value of 3000\,K.
 \label{fig_maps1a}} 
\end{figure*}

\begin{figure*}
\centerline{\includegraphics[angle=0,width=7.0in]{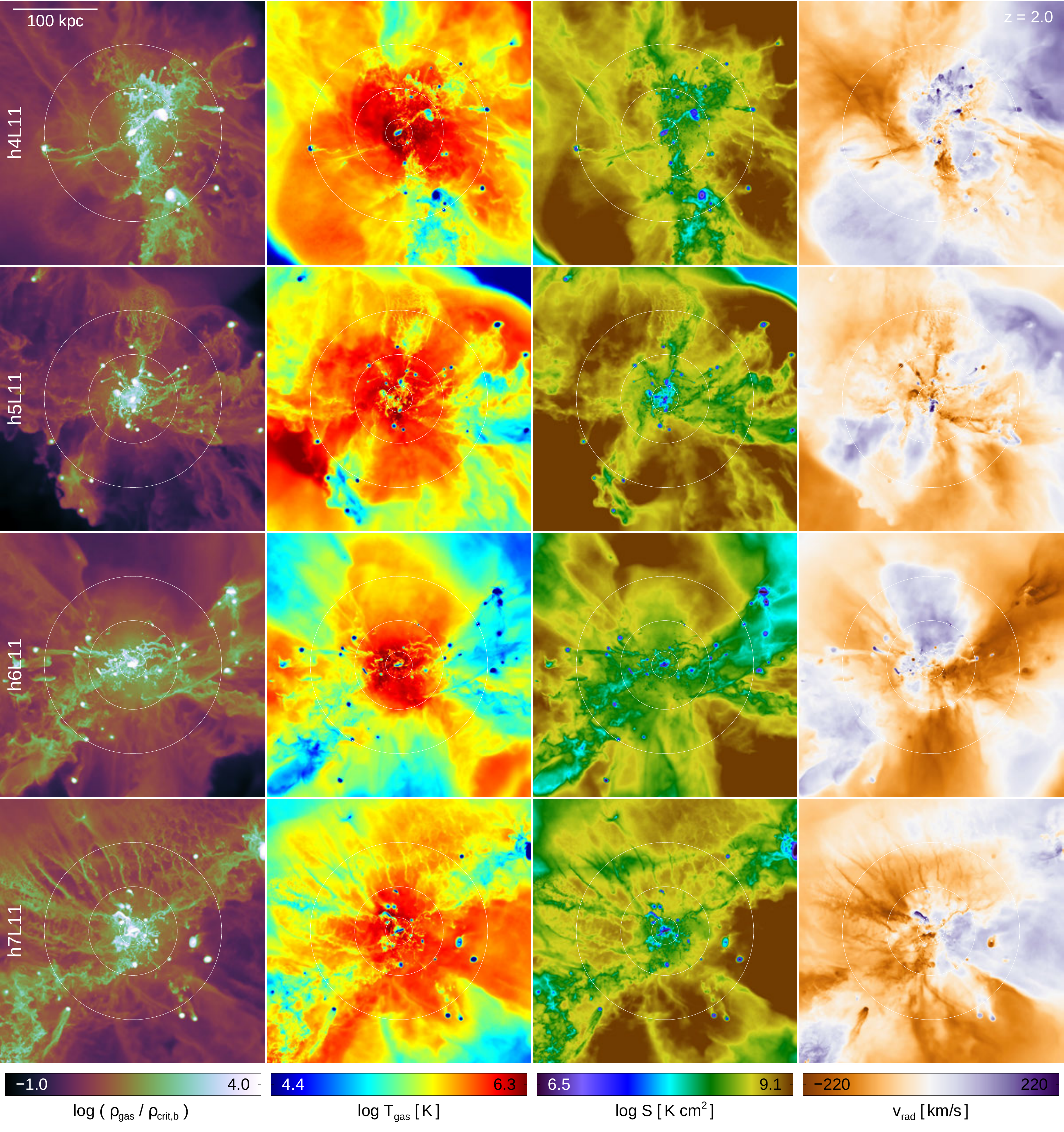}}
\caption{ As in the previous figure, mass-weighted projections of gas density, temperature, entropy, and radial 
velocity at $z\!=\!2$, but for the other four simulated haloes.
 \label{fig_maps2a}} 
\end{figure*}

We present a visual overview of the eight simulated haloes at $z\!=\!2$, focusing on the virial scale of haloes 0-3 
in Figure \ref{fig_maps1a}, and of haloes 4-7 in Figure \ref{fig_maps2a}, one halo per row. In each case, we 
show mass-weighted projections of gas density, temperature, entropy, and radial velocity in each column (left to right). 
All gas cells within a cube of side-length 3\,$r_{\rm vir}$ are included, while the temperature of star forming gas on 
the effective equation of state (eEOS) is set to a constant value of 3000\,K. The three concentric circles denote 
$\{0.15,0.5,1.0\} \times r_{\rm vir}$.

We have separated the eight systems into two groups based roughly on the morphology of their halo gas -- the first four 
haloes (h0-3) are closer to equilibrium at $z\!=\!2$ and the structure of their hot gas is roughly spherical, whereas the 
second four (h4-7) are significantly more disturbed. In the first case, the virial shock can be clearly seen as a sharp 
increase to higher temperature and entropy, typically at $\ga$\,1.25$r_{\rm vir}$. However, the radius of the shock 
\citep[see also][]{schaal15} or its existence at all depends strongly on direction, which we explore further in Section 
\ref{sAngVar}. These boundaries are also associated with a decreased inflow velocity -- that is, radial stalling -- as 
well as increased gas density. 

At $z\!=\!2$ haloes at this mass scale of $\simeq$\,$10^{12}$\,\msun typically reside at the intersection of multiple 
large-scale filaments (and/or sheets) of the cosmic web. This naturally leads to a significant amount of filamentary inflow across the 
virial radius. We see that these inflows cover only a fraction of the virial sphere. They can maintain coherency to at 
least half the virial radius while maintaining a strong overdensity with respect to the mean gas density at each radius. 
They experience significant heating, typically at $\sim$\,0.5$r_{\rm vir}$, from $\la$\,$10^{4.5}$\,K to $\ga$\,$10^6$\,K, 
reaching the peak temperature of the mean $T_{\rm gas}(r)$ in the inner halo. Their entropy can remain lower than that 
of the hot halo gas at these radii, $\sim$\,$10^{8.0-8.5}$\,K\,cm$^2$ compared to $\sim$\,$10^9$\,K\,cm$^2$. Although the 
mean radial velocity inside the halo is generally near equilibrium/zero (denoted by white in the colormap), we see that 
in a given direction this is rarely the case, finding instead either non-negligible inflow or outflow velocities.

Two haloes, h3 and h7, have a large number of small, prominent, inflowing streams in the radial range 
$0.5 r_{\rm vir} \la r \la 1.25 r_{\rm vir}$ (see density maps). They originate from the lower left direction in the case of h3, and from 
the top for h7. These filaments are not associated with large-scale structure, and have characteristics distinct from the 
much larger filaments which are associated with features at larger radii. In particular, although they obtain similarly 
high inflow velocity, their entropy and temperature are above that of the IGM, and their overdensity with respect to the 
radial mean is not as significant. Some of the other six haloes exhibit similar features at various points in the past, 
but they have disappeared by $z\!=\!2$. In general, we observe that these features form between one and two times the 
virial radius semi-spontaneously, as in an instability, commonly triggered by perturbations either from substructure 
debris or the intersection of sheet-like structure in the cosmic web. They are marginally evident at L10 and absent at 
L9. We reserve an in-depth analysis of the formation of these features and whether or not their growth corresponds to a 
physical instability in the gas for future work.

\begin{figure*}
\centerline{\includegraphics[angle=0,width=7.0in]{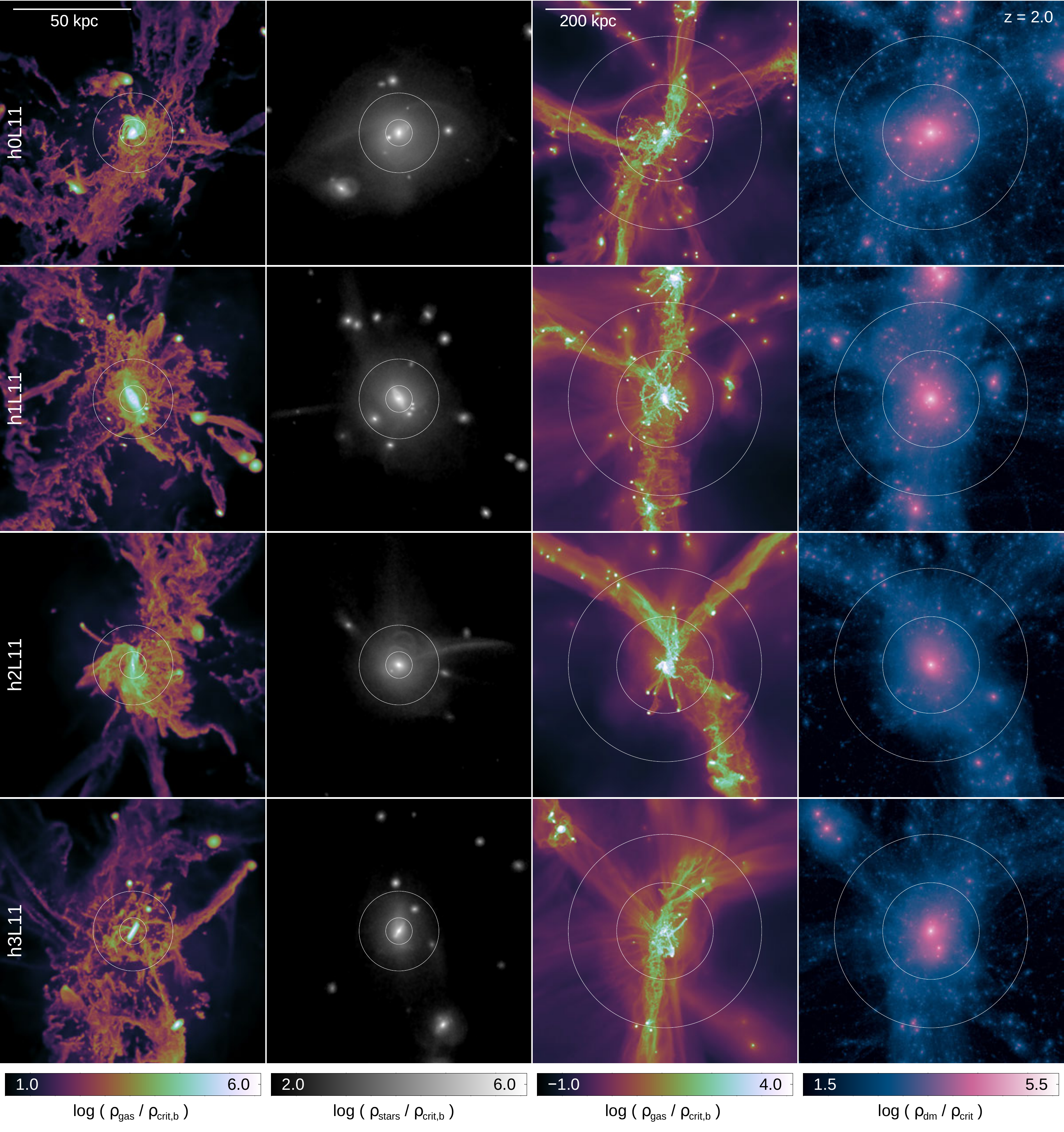}}
\caption{ Mass-weighted projections of gas and stellar density temperature at small scales (left two columns), and 
projected gas density and dark matter density at large scales (right two columns). The first four simulated haloes are 
shown at $z\!=\!2$. All gas cells or particles within a cube of side-length 1.0\,$r_{\rm vir}$ (small scales) or 
5.5\,$r_{\rm vir}$ (large scales) are included, and distributed using the standard cubic spline kernel with 
$h=2.5\,r_{\rm cell}$ (gas) or $h=r_{\rm 32,ngb}$ (the radius of the sphere containing the 32 nearest neighbours of the 
same particle type, for stars and dm) in orthographic projection. The white circles denote \{0.05,0.15\}\,$r_{\rm vir}$ 
(left two columns) and \{1,2\}\,$r_{\rm vir}$ (right two columns). Densities are normalized to the critical (baryon) 
density at $z\!=\!2$. 
 \label{fig_maps1b}} 
\end{figure*}

\begin{figure*}
\centerline{\includegraphics[angle=0,width=7.0in]{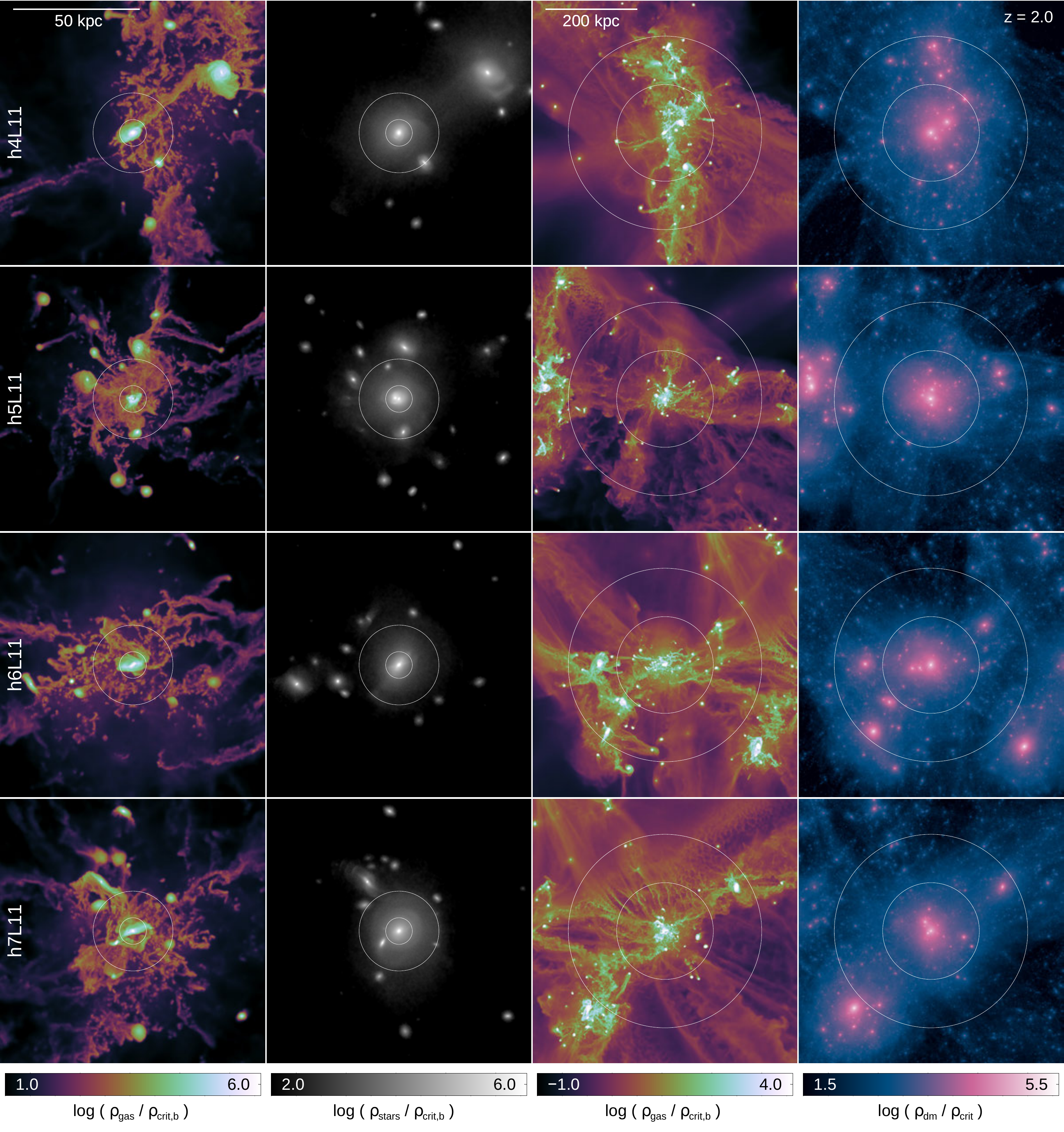}}
\caption{ As in the previous figure, projections of gas and stellar density on small scales (left two columns), 
and gas density and dark matter density on large scales (right two columns) at $z\!=\!2$, but for the other four 
simulated haloes. 
 \label{fig_maps2b}} 
\end{figure*}

We visualize two additional scales, the inner halo structure as well as the large-scale context, of the eight simulated 
haloes at $z\!=\!2$. Figure \ref{fig_maps1b} includes haloes 0-3 while Figure \ref{fig_maps2b} includes haloes 4-7. 
The left two columns show the projected density of the gas and stellar components, the white circles 
indicating 0.05\,$r_{\rm vir}$ and 0.15\,$r_{\rm vir}$. The right two columns show the projected density of the gas and 
dark matter components, the white circles here indicating 1.0\,$r_{\rm vir}$ and 2.0\,$r_{\rm vir}$. 

In the centre of the halo, $r/r_{\rm vir} \la 0.3$, the gas component forms a complex morphology with multiple orbiting 
substructures, commonly associated with tails and gas streams from ram-pressure stripping, as well as tidally-induced 
bridges and spiral patterns. Rapidly varying velocity fields modulate inflowing as well as outflowing material. At 
$r/r_{\rm vir} \la 0.05$ a large gas disc is a common feature -- the final stage in the transformation from 
large-scale structure and the associated tidal torques to the formation of a galaxy. Here the coherent angular momentum 
of the halo as a whole must transition to the angular momentum of the forming disc through the complicated dynamics of 
this inner halo `messy' region \citep{danovich14}. In the small-scale views of h2 we see tidal features in the 
collisionless stellar distribution (extending to the right from the centre), without any corresponding structure in the 
gas. In h3, we see a massive tidal tail in gas extending out to $\simeq$\,0.5$r_{\rm vir}$ (towards upper right of panel), 
with no corresponding structure in the stars. A recent post-pericenter passage of an ongoing major merger in h4 has 
induced tidal features in the large companion (upper right of figure) which is still on an outgoing orbit, and generated 
a long tidal tail, of which the inner half has reversed direction and is falling back on to the central galaxy. A recent 
merger in h5 has left a double nucleus in the stars, separating by roughly one kpc. In h7 we see a strong tidal warp in 
the gas and stars of the merging companion (to the upper left of the central), with several stellar shells from earlier 
minor mergers superimposed. In general, we can see that many of the narrow inflowing streams in the inner halo are tidal 
in nature, which highlights the importance of characterizing the origin of gas accretion features -- this component of 
the accretion rate would have been characterized as a `stripped' component in the methodology of \cite{nelson15a}.

In the inner regions of the halos which we have identified as in equilibrium, the perturbations and gas structures generated by the 
dynamics of satellite galaxies \citep[see][]{zavala12} are superimposed on a gas background which is aligned at Mpc scales with large-scale gas 
features. For example, the broad inflow in h0 (towards the top of the image panel) with a width of $\sim$\,50\,kpc at 
0.5\,$r_{\rm vir}$ is clearly associated with a gas overdensity extending out to at least 3\,$r_{\rm vir}$. Likewise 
for the broad inflow in h2 (also, top of the image panel) which has coalesced at the intersection of two such cosmic 
web filaments. These inflows are not smooth features of constant density gas. They have a `wavy' or rippled appearance, 
arising from gas density contrasts as high as a factor of $\sim$\,$10^3$. By visual inspection it is clear that, at least 
at sufficient resolution, such inflowing filaments are in fact associations of multiple smaller structures with coherent 
kinematics and origin.

The second set of haloes (h4-7), which are in general more morphologically disturbed, do not show this clear association 
between small-scale and large-scale gas features. Their time evolution indicates that this is often a consequence 
of a major halo-halo merger where the two gas reservoirs collide and disrupt any radially coherent gas kinematics as 
the hot halo re-equilibrates. However, recent major mergers are not always seen, and the nearby environment and topology 
of large-scale structure also appear to play a significant role in the sphericity of halo gas, independent of assembly 
history. For example, we consider the number of major mergers with ratio $\eta$\,$\ge$\,$1/3$ experienced by each halo 
while the stellar mass is greater than $10^{10.5}$\,\msun (roughly $2\!<\!z\!<\!4$), as well as the maximum merger ratio
of any merger, under the same constraint. In both cases, the mass ratio is determined at the time the less massive 
progenitor reaches its maximum stellar mass \citep[see][]{rodrig15}. We find that h0, h2, and h3 have had no major 
mergers exceeding this mass ratio, and indeed none with $\eta$\,$\ge$\,$0.2$. The outlier is h1, which did have one 
major merger ($\eta$\,=\,$0.8$). In the disturbed group, h4 has had four major mergers ($\eta_{\rm max}$\,=\,$0.6$), 
h6 has had one ($\eta$\,=\,$0.65$), and h7 has had three ($\eta_{\rm max}$\,=\,$0.95$). The outlier is h5, which has 
had no major mergers satisfying this criterion, the largest mass ratio being $\eta$\,=\,$0.2$. 



\section{The Physical State of Halo Gas} \label{sPhys}

In this Section, we quantify the gas behaviours noted above, first by giving the spherically-averaged profiles of gas 
density, cell size, temperature, entropy, radial velocity, and specific angular momentum, and then by examining the 
pairwise correlation matrix of these same gas quantities.

\subsection{Characteristic Halo Properties} \label{ssChar}

In what follows we will normalize the distance of gas cells from the halo centre by the 
virial radius, as calculated for each halo at each resolution. For simplicity, we will in general not normalize the 
thermodynamic properties of gas since our halo mass range is so narrow. Instead, to provide a sense of reference, we 
calculate representative values for gas temperature, entropy, density, velocity, and angular momentum for a halo of 
$M_{\rm halo} \simeq 10^{12}$\,\msun at $z\!=\!2$. Numerical values are given here. First, we take the virial temperature 
as

\begin{equation}
T_{\rm vir} = 
  \frac{\mu m_p v_{\rm vir}^2}{2 k_B} 
  \simeq 10^{6.3} \,{\rm K}
\end{equation}

\noindent following \cite{bl01}, where $\mu \simeq 0.6$ for a fully ionized, primordial gas. 
We take entropy as $S = P / \rho^\gamma$ where pressure is $P = (\gamma-1) T \rho$. A virial entropy is then defined as

\begin{equation}
S_{\rm vir} = \frac{P_{\rm vir}}{\rho_{vir}^\gamma}
            = (\gamma-1) T_{\rm vir} \rho_{\rm vir}^{(1-\gamma)}
            \simeq 10^{8.0} \,{\rm K \,cm^2}
\end{equation}

\noindent where we have taken $\rho_{\rm vir}=200 (\Omega_b / \Omega_m) \rho_{\rm crit}(z)$, the baryon fraction 
multiplied by two hundred times the critical density of the universe at that redshift. Using the chosen overdensity 
criterion, the virial radius is then

\begin{equation}
r_{\rm vir} = \left( \frac{G M_{\rm halo}}{100 H(z)^2} \right)^{1/3}
            \simeq 100 \,{\rm kpc}
\end{equation}

\noindent in physical units, while the virial velocity is

\begin{equation}
v_{\rm vir} = \sqrt{ \frac{ G M_{\rm halo} }{r_{\rm vir}} }
            \simeq 205 \,{\rm km/s}.
\end{equation}

\noindent Finally, for the specific angular momentum of the halo we take

\begin{equation}
j_{\rm vir} = \sqrt{2} \lambda v_{\rm vir} r_{\rm vir}
            \simeq 10^{3.0} \,{\rm kpc \,km/s}
\end{equation}

\noindent using a spin parameter of $\lambda = 0.035$ \citep{barnes87,bullock01}. These characteristic 
values for the temperature, density, entropy, radial velocity, and specific angular momentum of the halo gas are 
typically reached at some reasonable fraction of the virial radius, e.g. $\sim$\,$(0.1-0.5)$\,$\times$\,$r_{\rm vir}$. 
We use them below to provide a useful reference with which to identify halo gas in different physical states.

\subsection{Radial Gas Profiles} \label{ssRadial}

\begin{figure*}
\centerline{\includegraphics[angle=0,width=6.5in]{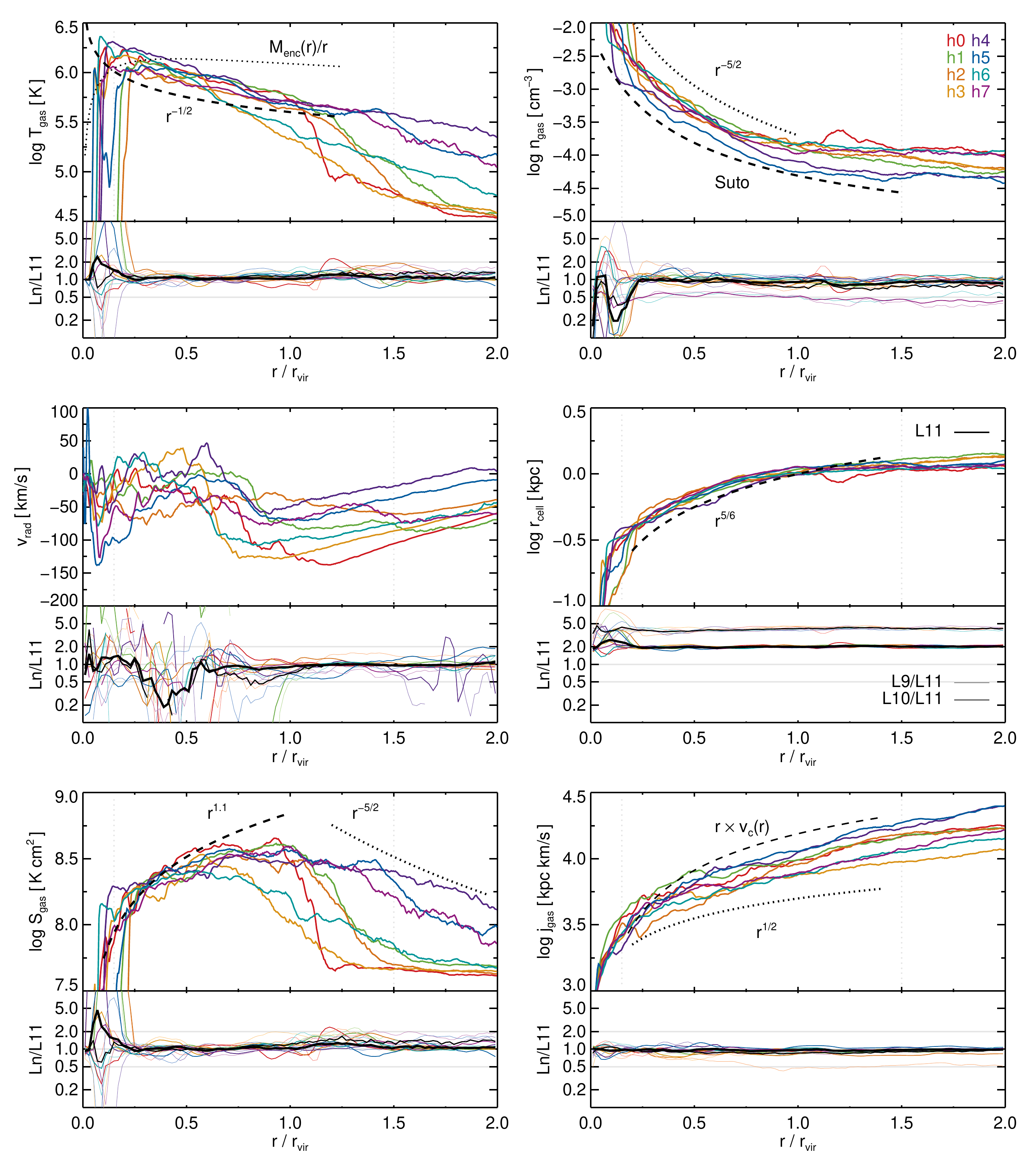}}
\caption{ The median radial profiles of six quantities: gas temperature, density, radial velocity, cell size, entropy, 
and angular momentum. Each halo is shown separately (by colour). The main panels for each quantity show the highest 
resolution only, while the subpanels show the ratio of the two lower resolution runs to L11 (the mean ratios across all 
eight haloes are indicated with thick black lines). Various scalings with radius are provided for reference (dotted and 
dashed lines). All gas within 2$r_{\rm vir}$ is included, except that substructures are excised.
 \label{fig_radial_profiles}} 
\end{figure*}

In Figure \ref{fig_radial_profiles} we show the median radial profiles of six gas quantities: temperature, density, radial 
velocity, cell size, entropy, and specific angular momentum. The main panel plots the profile for each of the eight 
haloes individually (different line colours) at the L11 resolution level. Light colours are used to indicate the four 
`equilibrium' haloes, while darker colours indicate the four more `disturbed' systems. The subpanels show the ratio of the 
radial profiles comparing the L9/L11 and L10/L11 runs, where the thick black lines indicate the average across haloes. 
We have excised all gravitationally bound substructures before constructing mean radial profiles, which would otherwise 
appear as a forest of small, dense, cold systems in these figures. A few observationally and theoretically motivated 
scalings with radius are included for reference (dotted and dashed lines). 

For instance, going beyond an isothermal model for gas temperature, we can take a gas component tracing dark matter 
distributed according to a NFW profile. This leads to a scaling of $T_{\rm gas} \propto M_{\rm enc}(r)/r$ 
\citep[e.g.][]{makino98}, which is still much too shallow (dotted line). Instead for $r/r_{\rm vir} \ga 0.15$ we 
have approximately $T_{\rm gas} \propto r^{-1/2}$. Of all eight systems, h0 exhibits the sharpest temperature drop 
defining an outer boundary of the halo at $r \simeq 1.2 r_{\rm vir}$, while in most cases there is no sharp transition 
at or near the virial radius. In all cases the temperature drops precipitously in the inner halo, at 5\%-10\% 
of $r_{\rm vir}$. Note that we have assigned a gas temperature of 3000\,K to gas on the star-forming equation of state, 
which would otherwise have an artificially high effective temperature. The $T_{\rm gas}(r)$ profiles are well converged 
with resolution. The mean across all eight haloes and averaged over $0.15 < r/r_{\rm vir} < 1.5$ is L9/L11\,$\simeq$\,1.15 
and L10/L11\,$\simeq$\,1.08, while at the largest radii the lower resolution runs have mean temperatures biased high by up 
to $\simeq$\,30\%. Physics will dominate resolution -- different models for galactic-scale stellar winds or AGN feedback 
can substantially modify the mean gas temperature profile inside as well as outside $r_{\rm vir}$ \citep{suresh15}.

A gas density scaling as $n_{\rm gas} \propto r^{-5/2}$ approximates the mean behaviour of the haloes well, as does a 
model for an isothermal gas in hydrostatic equilibrium \citep[][using $B=(n+1) B_p$ with $n=7$, $B_p=1$]{suto98}. However, 
the other panels make it clear that the gas is not in hydrostatic equilibrium, nor isothermal, implying this agreement 
may be coincidental. Convergence at lower resolutions is reasonable -- over the same radial range we find 
L9/L11\,$\simeq$\,0.8 and L10/L11\,$\simeq$\,0.9, 
indicating that the mean gas densities are lower in the lower resolution runs. This is potentially a consequence of 
better resolving satellite galaxies and their interactions with the central host, which fills the halo volume with 
more high density gas cells. These are no longer instantaneously bound to the satellite, and so are not 
excluded. \footnote{The filamentary feature in the upper right panel of Figure \ref{fig_mesh_slice_res}, extending from 
$\simeq$\,0.15\,$r_{\rm vir}$ to $\simeq$\,0.5\,$r_{\rm vir}$ is an example of such a tidal tail.} The total halo gas 
mass actually decreases somewhat at higher resolution ($\sim$\,10\% or $\sim$\,5\% lower for L11 as compared to L9 or 
L10, respectively), which would otherwise modify gas densities in the opposite direction. Finally, the variable assembly 
histories lead to the large halo-to-halo scatter, while the sensitivity of merger states to temporal offsets driven by 
short dynamical time-scales leads to large scatter between resolution levels. As with temperature, however, the caveat 
is that radial density profiles will depend in detail on the implemented feedback models \citep{hummels13}.

\begin{figure*}
\centerline{\includegraphics[angle=0,width=7.0in]{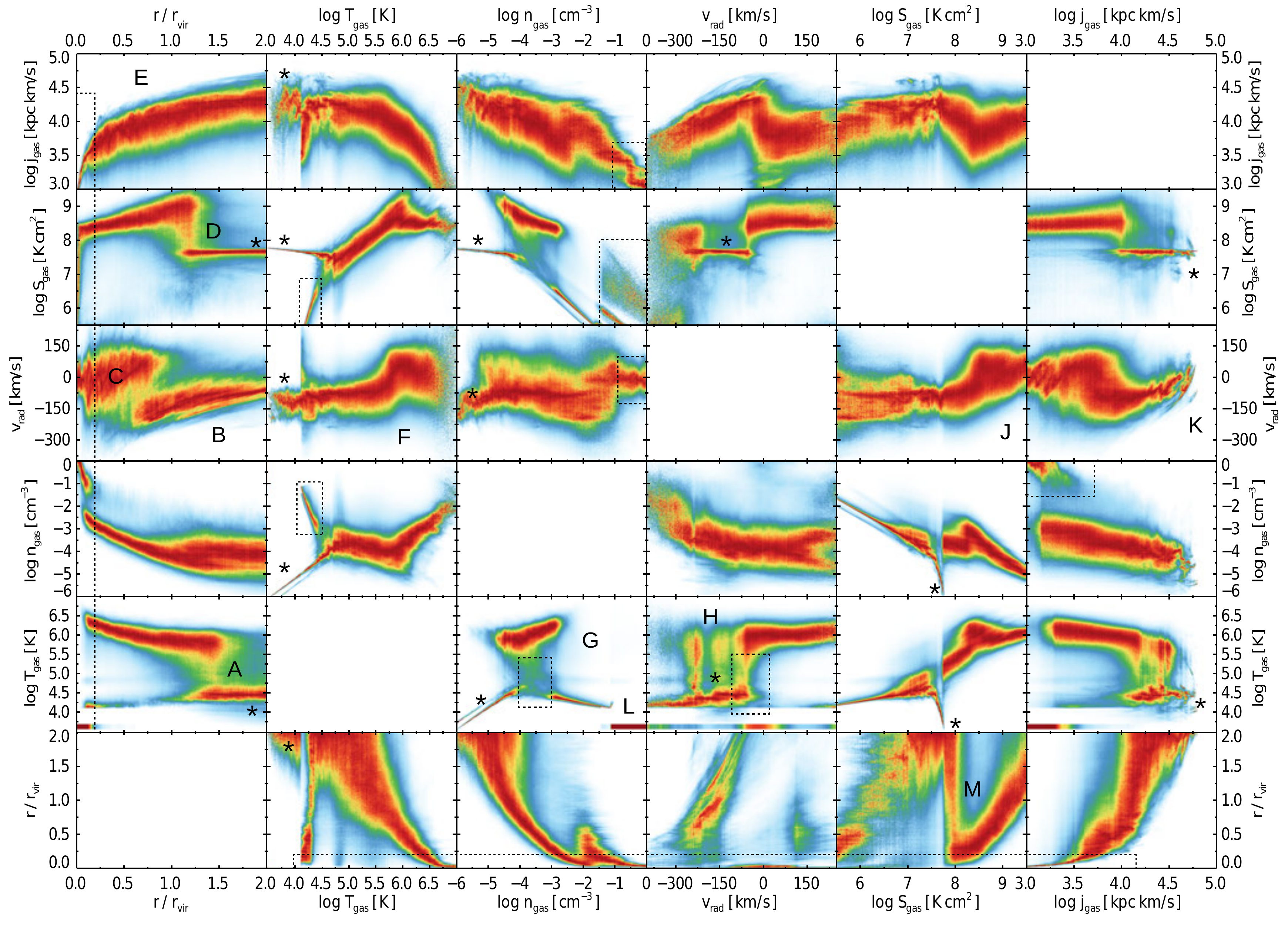}}
\caption{ The mass-weighted, two dimensional correlation matrix between all pairwise combinations of six quantities: gas 
radius, temperature, density, radial velocity, entropy, and angular momentum. All eight halos are included and stacked. 
Each radial bin is normalized independently, such that the colour mapping reaches its maximum intensity at each x-value, 
independent of the radial distribution of gas mass. Therefore, panels symmetrically across the diagonal from one another 
are not simply transpositions, but include additional information. Substructures are excised.
 \label{fig_matrix}} 
\end{figure*}

We see a similar scatter in the Hubble-corrected gas radial velocities. The haloes have different velocity structures, 
which commonly feature a slowly increasing inflow velocity down to $\sim$\,0.75\,$r_{\rm vir}$, at which point there is 
a noticeable bump towards an equilibrium value of $v_{\rm rad} \simeq 0$\,km/s. Until reaching the central galaxy, the 
radial velocity profiles are then roughly flat. In the following discussion of Figure \ref{fig_matrix}, however, we show 
that the mean (or median) velocity profile is an exceedingly poor representation of the dynamics of halo gas. For 
example, in the mean the quasi-static halo rarely has $v_{\rm rad}>0$ for $0.15 < r/r_{\rm vir} < 1.0$, while in 
actuality this value is driven down by the superposition of rapidly inflowing and gently outflowing components. 

The gas spatial resolution, for which we again use the sphere-equivalent radii $r_{\rm cell}$ as a proxy, becomes better 
with decreasing radius as expected. Given our constant cell mass refinement criterion, it scales roughly as the cube-root 
of the density scaling. That is, 
$r_{\rm cell} \propto V_{\rm cell}^{1/3} = (m_{\rm cell}/\rho_{\rm cell})^{1/3} \propto \rho_{\rm cell}^{-1/3}$. 
At L9 and L10 the cell sizes are a factor of four and two larger, respectively, scaling with the mean 
inter-cell spacing as $r_{\rm cell} \propto L_{\rm box} N_{\rm cell,tot}^{-1/3}$.

We overplot the gas entropy profiles with the near-linear scaling $S_{\rm gas} \propto r^{1.1}$ often seen in x-ray 
observations of local clusters \citep[e.g.][]{george09}. Despite looking at the significantly more massive cluster 
PKS 0745-191, $\sim$\,$10^{14}$\,\msun in the nearby universe ($z\,\simeq\,0.1$), the inferred radial characteristics 
from \cite{george09} are in reasonable agreement with the haloes simulated in this work. A similar level of qualitative 
agreement is evident at $z\!=\!2$ for $\sim$\,10$^{12}$\,\msun haloes in gas temperature, entropy, and density with 
previous cosmological simulations at L8-equivalent resolution and incorporating more realistic feedback physics 
\citep{vdv12b}. In both cases, entropy rises to near the virial radius at which point it flattens and becomes roughly 
constant. While some haloes exhibit this gradual plateau (h4), the radial entropy profile of others drops as $r^{-2}$ or 
faster (h2) at some radius $\ge r_{\rm vir}$. There is a correspondence between the equilibrium state of the halo and its 
properties beyond the virial radius -- the more disturbed systems (h4-7, darker line colours) generally have higher 
entropy, without any strong transition. The entropy profiles, averaged over all eight haloes, are well converged with 
respect to L9 and L10. \footnote{In the inner halo, h1 and h2 have sharp drops in both $T_{\rm gas}$ and 
$S_{\rm gas}$ at a larger radius than typical, $\sim$\,0.25$r_{\rm vir}$. In these two cases, while the profiles are 
properly centred on the galaxy, the galaxy is not entirely centred within the halo. This leads to more cold, low entropy 
gas beyond the typical disc size.} 

The specific angular momentum scales roughly as $r^{1/2}$ or even more accurately as $j_{\rm gas} \propto r\,v_c(r)$ 
given the circular velocity $v_c(r) \propto ( M_{\rm enc}(r)/r )^{1/2}$ for an NFW profile of this halo mass. There is 
broad uniformity among the eight haloes and good convergence in the lower resolution runs, with a small systematic bias 
towards lower angular momentum content in the halo, L9/L11\,$\simeq$\,0.88 and L10/L11\,$\simeq$\,0.96, again driven by 
high velocity tidal debris becoming better resolved at the highest resolution.

\subsection{Distributions Beyond Radial Dependence}

To avoid missing important features by averaging over multiple gas populations with distinct properties residing at similar radii, 
we examine the two dimensional distributions of these same gas quantities. Instead of only considering their radial 
dependence, in Figure \ref{fig_matrix} we show a full pairwise correlation matrix, every quantity plotted with respect 
to every other. Here we have stacked together all eight haloes at L11, normalizing only the radius of each gas cell with 
respect to $r_{\rm vir}$ of its parent halo. Despite the different thermal and dynamical structures of the haloes we 
verify that the key features seen individually in each are preserved in this collective view. To explore the various 
relations, we have separately normalized each panel `column by column`. That is, for each value along the x-axis, the 
colour mapping of the corresponding one dimensional, vertical slice is independent, such that each slice extends from 
low values (white/blue) through intermediate values (green/yellow) to high values (orange/red). Therefore, there is no 
global indication of the distribution of mass within each panel (although very noisy regions indicate a poor sampling 
and therefore comparably little gas mass). In addition, the corresponding panels across the diagonal are not just 
transpositions of one another, but answer different questions. For example, the $T_{\rm gas}(r)$ panel in the first 
column indicates, for each radius, the dominant temperature(s) of gas at that radius, whereas the $r(T_{\rm gas})$ panel 
in the second column indicates, for each temperature, the dominant radius (or radii) of gas with that temperature. We 
describe some of the more interesting features seen in this matrix, which have been labelled with letters (A)-(M).

\begin{itemize}
\item (*) Denotes features which arise in the intergalactic medium and disappear if gas at $r > r_{\rm vir}$ is 
excluded. In particular, all the strong constant entropy features at $S_{\rm gas}$\,$\simeq$\,$10^{7.7}$\,K\,cm$^2$ in 
the second row from the top, and the near-discontinuities at this same value in the second column from the right.

\item \framebox{\phantom{t}} Regions demarcated with dotted rectangles indicate features arising in the galactic disc or 
its vicinity, and disappear if gas at $r < 0.2 r_{\rm vir}$ is excluded. This includes all low $j$ gas at high density, 
which is already above the threshold for star formation in the galaxy. It also includes low temperature gas which has a 
radial velocity near zero, as well as the thin horizontal feature of $v_{\rm rad}$ spanning $\sim$\,0\,$\pm$\,100\,km/s 
near $r=0$ corresponding to a rotating disc. Interestingly, the prominent temperature bridge between the hot halo and 
galaxy at intermediate densities also largely disappears, indicating that a majority of this cooling occurs at 
$r < 0.2 r_{\rm vir}$ (see also G).

\item (A) The virial shock seen as a sharp temperature jump from $T_{\rm IGM} \sim 10^{4.5}$\,K to $\sim$\,$10^6$\,K. 
The width of this transition varies greatly between the eight haloes, h0 being the narrowest at 
$\Delta r/r_{\rm vir} \simeq 0.1$, and as seen here in the stack broadened to $\Delta r/r_{\rm vir} \simeq 0.5$ or more. 
The radius where this transition occurs is typically between $1.25$\,$r_{\rm vir}$ and $1.5$\,$r_{\rm vir}$. After 
shocking, the temperature slowly increases with radius, following the mean radial profile of the halo, until it reaches 
$\sim$\,0.25$r_{\rm vir}$ where higher densities lead to accelerated radiative cooling, allowing gas to join the cold 
ISM phase.

\item (B) The radial velocity profile of inflowing gas. This component speeds up as it flows in from large distances 
down to $\sim$\,0.5$r_{\rm vir}$ at which point it largely disappears. The mean velocity is approximately -75\,km/s at 
2\,$r_{\rm vir}$, -150\,km/s at $r_{\rm vir}$, and -225\,km/s at $r_{\rm vir}/2$. For free-fall from rest at infinity to 
a point mass of $10^{12}$\,\msun we would expect a speed of -200\,km/s at twice the virial radius. The actual value is 
less due the combination of the Hubble expansion and gas dynamics. Given this offset at the halo outskirts, the 
subsequent scaling is roughly consistent with $v_{\rm rad} \propto r^{-1/2}$ as expected from free-fall, at least down 
to $\sim$\,0.5\,$r_{\rm vir}$, below which inflow no longer dominates by mass.

\item (C) The radial velocity distribution of the hot halo gas. These two components overlap between 
$0.5 < r/r_{\rm vir} < 1.0$ as gas transitions from rapidly inflowing to quasi-static. It really is `quasi', however, 
since a large mass of gas has positive radial velocity. Gas with $v_{\rm rad} > 0$ arises from the dynamical formation of 
the halo atmosphere and associated splashback motion in the baryons \citep[e.g.][]{wetzel15}. In this stacked view the 
radius of this transition is notably interior to the virial shock, $\sim$\,0.75$r_{\rm vir}$ as compared to 
$\sim$\,1.25$r_{\rm vir}$. Investigating the haloes individually we conclude that this is radial offset is largely a 
misleading feature arising from the stacking of radially averaged velocity profiles. Instead, the radius of a strong 
jump in temperature and entropy also closely corresponds to a sudden decrease of inward velocity, as a result 
of the transfer of kinetic to thermal energy.

\item (D) The virial shock seen in a sharp entropy jump from the nearly constant value in the IGM of 
$\sim$\,$10^{7.7}$\,K\,cm$^2$ to $\sim$\,$10^9$\,K\,cm$^2$ characteristic of the hot gas for haloes of $10^{12}$\,\msun. 
The entropy increase occurs over a narrower radial range than the corresponding temperature increase, and at a slightly 
smaller radius with respect $r_{\rm vir}$. This may be indicative of pre-shock compressive heating, although we caution 
that this radial difference varies from halo to halo and as shown in the stack. For example, in our subsequent exploration 
of h0 in Figure \ref{fig_rays2dall} we see that there is a close correspondence between the radii of temperature and 
entropy jumps. After shocking, the entropy slowly declines with radius, following the mean radial profile of the halo.

\item (E) The angular momentum distribution is uni-modal and scatters about its mean profile, with no evidence for 
multiple populations of gas having distinct $j_{\rm gas}$ at any radius.

\item (F) The radial velocity depends somewhat on the temperature of the gas, with the hottest gas populating the 
high positive velocity tail, while cold gas is inflowing. At all temperatures there is a continuous distribution of 
radial velocities, with an upturn at $\sim$\,$10^6$\,K, above which gas has a mean radial velocity consistent with zero, 
and below which the mean $v_{\rm rad}$ is always negative \citep[see also][]{joung12}. The rapidly inflowing hot gas, 
discussed in (C), is the high velocity tail of the gas distribution at these temperatures and so sub-dominant by mass.

\item (G) The usual `phase-diagram' plotted for cosmological simulations. Cold, low density gas 
($n_{\rm gas} \la 10^{-4}$\,cm$^{-3}$) in the IGM (denoted by a star) occupies the lower left corner, the tight relation 
with $T \propto n^{2/3}$ indicative of adiabatic compression. Shock-heated gas at intermediate densities is the only 
source for $T > 10^{5.5}$\,K gas. At higher densities ($n_{\rm gas} \ga 10^{-3}$\,cm$^{-3}$) strong cooling flows 
develop at small radii, after which gas approaches the effective temperature floor of $\sim$\,$10^4$\,K until it reaches 
the star formation threshold at $n_{\rm H}=0.13$ cm$^{-3}$. 

\item (H) The heating of inflow, seen here as the dominant temperature for gas at each radial velocity. The handful of 
strong vertical features are largely an artefact of the stacking. For a given halo, when we restrict this panel to include 
only gas with $0.2 < r/r_{\rm vir} < 1.0$ we find a clear correlation between deceleration and heating, where gas transitions 
from cold at -300\,km/s to hot by -150\,km/s. This is decidedly interior to the virial shock, and outside the galaxy.

\item (J) The average radial velocity as a function of entropy, restricted to gas in $0.2 < r/r_{\rm vir} < 1.0$ (not 
shown) resolves nicely into two distinct populations, where material with $S_{\rm gas} \ga 10^{8.5}$\,K\,cm$^2$ has 
a mean $v_{\rm rad}=0$ which is roughly constant. On the other hand, gas with lower entropy than this threshold has 
a mean $v_{\rm rad} \simeq -150$\,km/s, with a gradual trend towards faster inflow with lower entropy.

\item (K) Similarly, the average radial velocity as a function of angular momentum, restricted to gas in 
$0.2 < r/r_{\rm vir} < 1.0$ resolves nicely into two components. Gas with $j \la 10^{4.25}$\,kpc\,km/s has zero mean 
radial velocity, whereas higher angular momentum gas has $v_{\rm rad} \simeq -v_{\rm vir}$ (this is not shown explicitly 
in the figure, which includes all gas). The two populations overlap for $10^{4.25} < j_{\rm gas} < 10^{4.5}$. Together 
with (J) this panel shows the distinct physical properties of the quasi-static and inflowing halo gas.

\item (L) Star forming gas has a temperature on the effective equation of state, which is set to a low, constant value 
to differentiate it from other gas. This results in the constant temperature band at $\simeq$10$^{3.5}$\,K across this 
entire row. This gas is restricted to small radii, high densities, low radial velocities, and low angular momenta.

\item (M) Gas with high entropy, $S \ga 10^8$\,K\,cm$^2$ has a well defined relation to radius, following the mean halo 
profile. Below this threshold, if we exclude material in the IGM, gas at any given entropy is essentially distributed 
throughout the entire halo.
\end{itemize}

We have seen in several cases how the mean or median radial profiles fail to fully capture the full state and structure 
of halo gas. Clear examples are the distributions of temperature, entropy, and radial velocity as a function of radius. 
On the other hand, the distributions of density and angular momentum as a function of radius are comparatively well 
described by a median and scatter. By measuring average properties within radial bins we implicitly assume that the 
structure of halo gas is spherically symmetric. Even Figure \ref{fig_matrix} fails in this regard. For instance, the 
finite radial thickness of the temperature jump associated with the virial shock could either arise from (i) a 
spherically symmetric feature of that same thickness, or (ii) the superposition of many thin shocks spread throughout 
the same radial range, different shock fronts existing at different radii depending on direction. Motivated by the 
latter possibility, we proceed to investigate and quantify any variation of the structure of halo gas as a function of 
angle on the sphere.


\section{Angular Variability} \label{sAngVar}

\begin{figure*}
\centerline{\includegraphics[angle=0,width=7.0in]{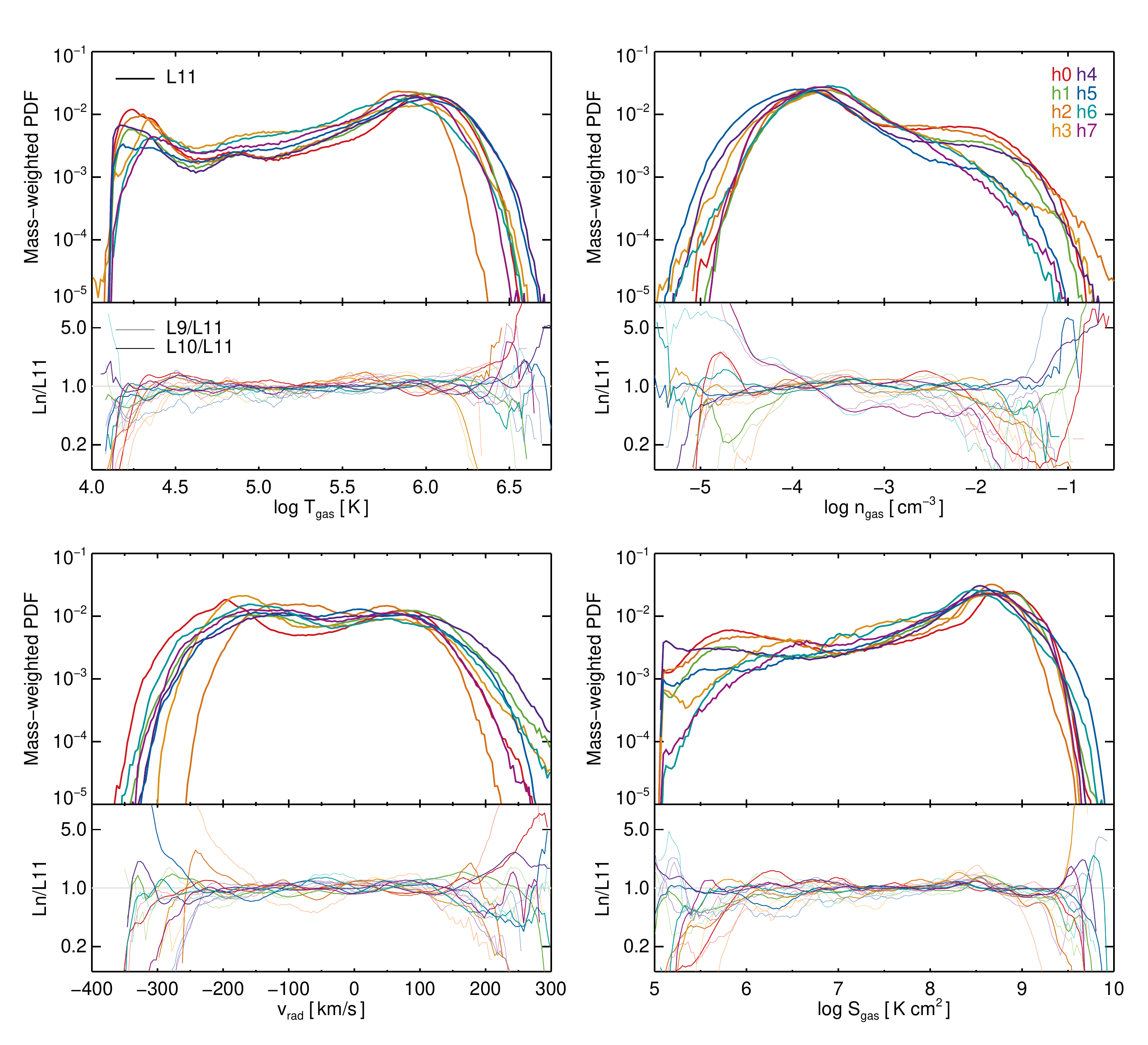}}
\caption{ The mass-weighted PDF of instantaneous gas temperature, density, radial velocity, and entropy, restricted to 
the radial range $0.5 r_{\rm vir} < r_{\rm gas} < 1.0 r_{\rm vir}$. The upper panel shows all eight haloes at the highest 
resolution level (L11, different colours), while the subpanels below show the ratio of the two lowest resolution levels 
to the highest, L9/L11 (thin lines) and L10/L11 (medium lines). The vertical axes of the ratio subpanels are 
logarithmic from 0.1 to 10. For the most part, the eight haloes show similar structure and their lower resolution 
counterparts scatter about the L11 distributions (see text for details).
 \label{fig_pdfs}} 
\end{figure*}

As a first step away from radial averages, we consider the properties of gas in a broad radial shell encompassing the 
regime of interaction between quasi-static material and filamentary inflow. In Figure \ref{fig_pdfs} we plot the 
distributions of temperature, density, radial velocity, and entropy, for all gas cells within 
$0.5 r_{\rm vir} < r_{\rm gas} < 1.0 r_{\rm vir}$. In particular, we examine how well the lower resolution runs 
reproduce the distributions of these quantities found in the highest resolution (L11) runs. In this radial range, we 
find that the gas of all simulated haloes has broadly similar properties, differing only in the details. 

The temperature distribution (upper left) shows a broad peak centred roughly at $T_{\rm vir}$, 
slowly falling off towards lower temperatures and with a distinct low temperature peak at $\simeq$\,$10^{4.25}$\,K.
While the fractional amount of gas at each temperature in the L9 and L10 runs can differ throughout this regime by up to 
a factor of two, depending on halo, the mean ratio of both L9/L11 and L10/L11 is consistent with unity from $10^{4.5}$\,K to  
$10^{6.5}$\,K. At the highest temperatures $>$\,$10^{6.5}$\,K the lower resolution runs have much larger deviations with 
respect to L11, but we attribute this primarily to poor sampling of the extreme tail of the distribution due only to a 
low number of available gas cells. In contrast, the PDF at the low temperature peak is systematically lower in the lower 
resolution runs. At $\simeq$\,$10^{4.25}$\,K where there are still a large number of gas cells, the mean across all eight 
haloes is L9/L11\,$\simeq$\,0.5 and L10/L11\,$\simeq$\,0.8, indicating that there is a smaller fraction of the total gas 
mass at these temperatures at lower numerical resolution. Note that $10^4$\,K is effectively the temperature floor due to 
cooling in these simulations, and the mean $T_{\rm IGM}$ at redshift two is just above this value.

Gas density shows an even larger variation on a halo to halo basis -- some have a strong second peak at higher densities, 
while this feature is largely absent for the less relaxed haloes. The ratio with respect to the lower resolution runs is 
consistent with unity for $\log \,n_{\rm gas} < -2.5$\,cm$^{-3}$. Between this value and the star formation threshold of 
$\simeq$\,$10^{-1.0}$\,cm$^{-3}$, there is a decrease, at lower resolutions, in the fractional amount of gas at these 
densities. In this range, the mean ratio across all eight haloes is L9/L11\,$\simeq$\,0.4 and L10/L11\,$\simeq$\,0.6. As 
with temperature, we see that the density PDFs can disagree between resolutions by as much as a factor of ten, but only 
in the tails when the magnitude drops to low values of $\leq 10^{-4}$. The disagreement is similar with radial velocity, 
where the mean Ln/L11 ratios are consistent with unity for all values away from the extremes.

The distribution of $v_{\rm rad}$ itself is comprised of two 
broad components, roughly centred about zero, the positive and negative peaks indicative of outflow and inflow, 
respectively. The primary driver of inter-halo variation -- for example, that h0 has smaller minimum and maximum 
velocities -- appears to be differences in assembly history, particularly a recent merger with another massive halo, or 
lack thereof. We explore this further in the following section. Finally, gas entropy behaves similarly to temperature, 
where we find that low entropy gas is strongly sub-dominant by mass with respect to the high entropy hot halo material.
The largest variation between haloes occurs for $\log \,S_{\rm gas} < 10^6$\,K\,cm$^2$, down to the floor at 
$\simeq 10^5$\,K\,cm$^2$. The halo-average PDF at L9 and L10 reveals a smaller fraction of gas at these lowest entropies 
when compared to the L11 run. 

In general, we conclude that for any given halo, large variations between the three resolution levels can be seen, with 
deviations up to a factor of ten in the fractional amount of gas at a particular temperature, density, radial velocity, 
or entropy. These deviations occur mostly in the tails of the distributions which are poorly sampled at the 
lower resolution levels. Furthermore, the timing differences present in any single comparison likely influence some of 
the largest outliers. With respect to the average behaviour across all eight haloes, we find that the lower resolution 
runs have less low temperature, high density, and low entropy gas. By mass fraction with respect to the total gas mass 
in this radial range, the magnitude of the effect is $\sim 2$ ($\sim 1.5$) for L9 (L10), although it is unclear if this 
gas population is necessarily converged for these haloes at L11. 

\subsection{Structure Along Radial Sightlines}

\begin{figure*}
\centerline{\includegraphics[angle=0,width=7.0in]{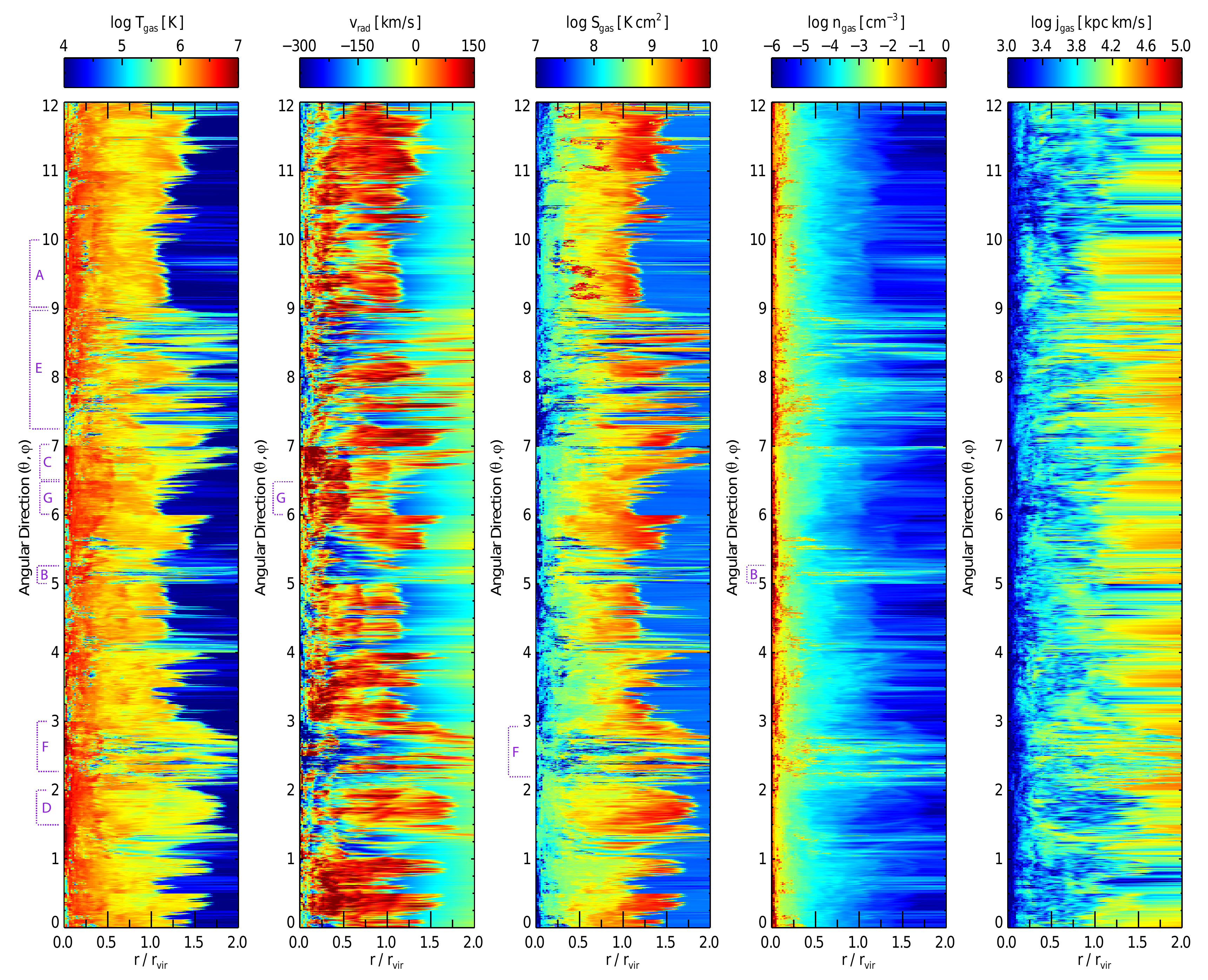}}
\caption{ The angular variability of gas temperature, radial velocity, entropy, density, and angular momentum at a given 
radius. Each pixel along the vertical direction represents a single radial ray, which are equally spaced in angular 
separation. The radial variability of the virial shock is evident in the shifting boundary between dark blue and 
yellow/orange in the left panel. Penetration of low temperature gas to radii smaller than $r_{\rm vir}$ correlates with 
higher inflow velocities, lower entropy, and higher density, often extending out into the IGM at $r \ge 2 r_{\rm vir}$.
In this figure we include only a single halo, h0L11, with $N_{\rm side}=16$, $N_{\rm rad}=100$. All substructures other 
than the primary have been excised, thereby excluding satellites.
 \label{fig_rays2dall}} 
\end{figure*}

To proceed, we consider sightlines originating outwards from the halo centre in different directions, which will have 
different radial structure. If the hot halo gas was triaxial, for example, we could anticipate that the radius of a strong 
virial shock would differ along its major and minor axes. We therefore measure the angular variability of the thermal and 
dynamical structure of the halo by casting many such sightlines from the centre of each halo. Along each of these 
`radial rays' we sample the continuous fields of gas temperature, radial velocity, entropy, density, and angular 
momentum in fixed steps of $\Delta r$. Using this ensemble of rays we can then quantify the structure of halo gas without 
binning in spherically symmetric radial shells.

Ray directions are set using the {\small HEALPIX} discretisation of the sphere \citep{gorski05} into equal area pixels, 
which implies equal angular spacing of rays at all refinement levels. The total number of radial rays is 
$N_{\rm ray} = 12 N_{\rm side}^2$, corresponding to an area subtended by each ray equal to  
$\Omega_{\rm ray} = 4 \pi / 12 N_{\rm side}^2$ sr, or $\theta_{\rm ray} = (180^2 / 3 \pi N_{\rm side}^2)^{1/2}$ deg. The 
sampling in the radial direction is linear and controlled by the parameter $N_{\rm rad}$ such that 
$\Delta r / r_{\rm vir} = 2.0 / N_{\rm rad}$. Our fiducial parameters of $N_{\rm rad}=400$, $N_{\rm side}=64$ result in 
a sampling of $\Delta r \simeq 0.5$\,kpc and $\Delta \theta = 1$\,deg. At each point, mean gas properties are estimated 
with a tophat kernel with adaptive size equal to the radius of the sphere enclosing the $N_{\rm ngb}=20$ nearest gas 
neighbours.\footnote{This is close to the mean number of natural neighbours (or cell faces) of the evolved Voronoi mesh in a 
cosmological simulation \citep{vog12}. Therefore the spatial scale of this effective smoothing is approximately matched 
to the scale of the stencil used in gradient estimation for linear reconstruction of fluid quantities in the MUSCL-Hancock 
scheme of {\small AREPO} \citep{spr10}.}

In Figure \ref{fig_rays2dall} we show a snapshot of the combined radial and angular structure of a single halo (h0L11) 
at $z\!=\!2$. The twelve demarcated intervals along the y-axis denote the twelve base pixels of the {\small HEALPIX} 
discretisation. Within each base pixel the nested ordering scheme uses a hierarchical quad-tree to preserve adjacency, 
and the four sub-intervals delineate the top nodes of each such quad-tree, implying that structure seen in the vertical 
direction is spatially coherent within these major and minor intervals. We note that there is no mass weighting or, for 
that matter, any indication of the mass distribution within each of the five panels, since the sampling points are 
smoothly distributed throughout the volume regardless of the underlying gas cell distribution.

Focusing first on the temperature structure we clearly see a boundary separating the cold, intergalactic medium from 
hot, virialised gas. At this halo mass scale, the temperature increases by roughly two orders of magnitude, from 
$\simeq$10$^4$\,K to $\simeq$10$^6$\,K (dark blue to yellow/orange). In some directions this heating is coherent and 
occurs at essentially uniform radius (for example, 9-10, marked `A'), demarcating a clear `virialization boundary'. However, 
across all sightlines we also observe a temperature jump of similar magnitude anywhere from $1.0 < r/r_{\rm vir} < 1.5$, 
and smaller jumps can occur out to twice the virial radius. Even so, from Figure \ref{fig_maps1a} we know that this 
halo is one of the most spherically symmetric, with a noticeable transition in the state of gas just outside the virial 
radius. In directions where the gas temperature is warm and exceeds $\sim$10$^5$\,K out to twice $r_{\rm vir}$, we find 
a correspondence to a baryonic overdensity, with respect to mean at that distance. This is indicative of large-scale gas 
filaments and the heating associated with their earlier collapse.
 
Interestingly, in directions where gas coherently penetrates to radii smaller than the virial radius and remains cold 
(e.g. B), this same correspondence to overdensities at larger distance remains. Qualitatively, the existence of an 
inflowing gas filament arising from the cosmic web suppresses a strong shock at the virialization boundary. This can 
arise from previous heating from filament formation outside of the halo, which increases the gas temperature 
to an intermediate state between $T_{\rm IGM}$ and $T_{\rm vir}$. There can nonetheless be a (smaller) temperature jump 
around the mean virialization boundary (C), although this is not always true and gas can gradually heat seemingly all 
the way to the peak of the halo temperature profile just exterior to the disc region (e.g. D).
Alternatively, this shock suppression can also arise from a delay of strong heating to deeper within the halo, in which 
case there is less preheating evident at large distances. The radius of the strongest temperature jump then decreases, 
typically to $0.25 r_{\rm vir} < r < 0.75 r_{\rm vir}$ as in much of (E). There are essentially no sightlines along 
which the gas does not experience heating above $T_{\rm IGM}$ at some radius. Since the velocity field, and so the 
streamlines of accreting gas, are not purely radial, this does not preclude the possibility that gas could avoid heating 
along its actual dynamical path. Such a trajectory would be curved in these panels. In some cases (F) we can largely 
rule out this behaviour, since over a large coherent solid angle no cold gas at $r < 0.5 r_{\rm vir}$ is directly 
connected to cold gas at larger radii. In general, however, the analysis based on radial rays is limited in this respect.

As with density, we also see a strong correlation between temperature and both entropy and radial velocity. Strong 
temperature jumps just outside the virial radius are also evident as sharp transitions from the gradually increasing 
(negative) IGM inflow velocity, through equilibrium, to a generally (positive) outflow speed indicative of the virialised 
halo gas. In some directions (G) multiple components are clearly visible, where an inner halo is comprised of 
hotter, denser, and more rapidly expanding gas which has more recently bounced back under its own thermal pressure. 
Similarly, the entropy of the low-density IGM undergoes a sharp increase before declining towards the halo centre. On 
a sightline by sightline basis the radius of the entropy jump is nearly equal to that of the temperature and radial 
velocity jumps. Finally, the angular momentum of the gas is clearly the least affected by the virialization boundary. 
As the velocity field becomes increasingly complex towards smaller radii, the non-radial component contributing to the 
angular momentum does not have a clear correlation with any of the other local gas properties. 

\subsection{Different Gas Heating Regimes}

\begin{figure*}
\centerline{\includegraphics[angle=0,width=7.0in]{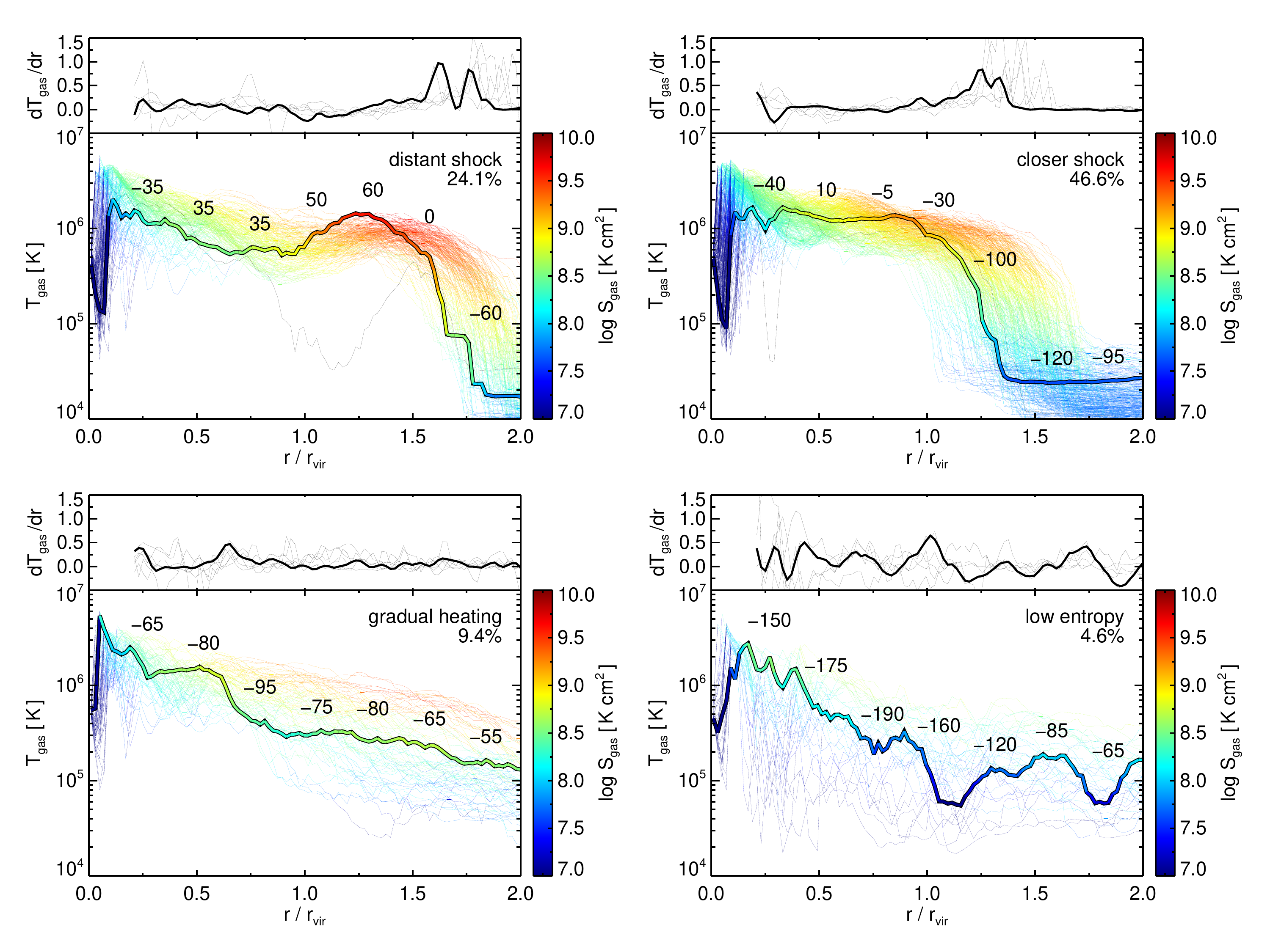}}
\caption{ The temperature profiles of individual radial rays (four main panels), with colour indicating gas entropy. 
Each of the four panels includes a disjoint subset of the entire ray set, where the selection for each was chosen by 
visual inspection in order to find four types of sightlines, each with similar radial properties, which together 
cover the majority of behaviours. In each panel, all rays are shown as thin lines, while a single prototypical example 
is shown as a thick coloured line. The mean radial velocity profile for each ray type is indicated by the series of 
numbers (in units of $km/s$, negative denoting inflow). The smaller top insets show the derivative of gas temperature 
with respect to radius, in units of (log K)/(0.1 $r_{\rm vir}$). The percentage in each panel indicates the fraction of 
all rays satisfying the criteria, for this particular halo. Given our selections, some sightlines are excluded from all 
four types. The vast majority of these rays either fall just outside a selection, or experience a significant temperature 
dip within the halo, which is a signature of satellite debris or intersection with non-radial filamentary inflows. Two 
such examples are shown as thin black lines in the upper two panels. In this figure we include only one halo, h0L9, 
with $N_{\rm side}=8$ and $N_{\rm rad}=100$ for visual clarity. 
 \label{fig_rayselect}} 
\end{figure*}

To understand the properties and the importance of sightlines exhibiting different radial behaviours, we would like to 
identify a few characteristic types. We separate rays into a small number of disjoint sets in Figure \ref{fig_rayselect}, 
each group ideally having a distinct radial behaviour. By visual classification we identify four such groups, which 
together encompass the majority of sightlines. In particular, we split rays into `distant shock', `closer shock', 
`gradual heating', and `low entropy' types, each of which is based on a quantitative selection applied across some 
radial range. Specifically, for the `distant shock' type we require all of 

\begin{itemize}
  \item {\rm max}(${\rm d}T/{\rm d}r |_{0.2 < r < 2.0}) > 0.25$
  \item $1.5 < r_{{\rm max}({\rm d}T/{\rm d}r)} < 2.0$
  \item ${\rm min}(T_{\rm gas}|_{0.2 < r < 1.4}) > 2 \times 10^5 \,{\rm K}$
\end{itemize}

\noindent where ${\rm d}T/{\rm d}r$ denotes the derivative of gas temperature with respect to radius, in units of 
(log K)/(0.1\,$r_{\rm vir}$), $r_{{\rm max}(Q)}$ indicates the radius where the quantity $Q$ reaches its maximum value, 
and $Q|_{r_1<r<r_2}$ indicates that a quantity $Q$ is constrained only within the radial range between $r_1$ and $r_2$. 
For the `closer shock' type we require

\begin{itemize}
  \item {\rm max}(${\rm d}T/{\rm d}r |_{0.2 < r < 2.0}) > 0.25$
  \item $0.8 < r_{{\rm max}({\rm d}T/{\rm d}r)} < 1.5$
  \item ${\rm min}(T_{\rm gas}|_{0.2 < r < 0.7}) > 3 \times 10^5 \,{\rm K}$.
\end{itemize}

\noindent That is, both must exhibit a strong temperature jump at either large or intermediate radii, while 
excluding rays which subsequently drop to low temperature at smaller radii -- an artefact of intersecting non-radial 
cold debris, as we subsequently discuss. Of the remaining rays not meeting the prior two conditions, we further 
require rays of a `gradual heating' type to satisfy

\begin{itemize}
  \item {\rm max}(${\rm d}T/{\rm d}r |_{0.2 < r < 2.0}) < 0.8$
  \item ${\rm min}(T_{\rm gas}|_{0.2 < r < 0.75}) > 3 \times 10^5 \,{\rm K}$.
\end{itemize}
  
\noindent Finally, the `low entropy' type requires of any remaining rays

\begin{itemize} 
  \item ${\rm max}(S_{\rm gas}|_{0.2 < r < 2.0}) < 8 \times 10^8 \,{\rm K \,cm^2}$.
\end{itemize}

\noindent In Figure \ref{fig_rayselect} we show all four groupings as separate panels (for halo h0L9). The temperature 
is plotted as a function of radius, while colour indicates gas entropy. A single prototypical ray is included as a thick 
line, while all rays belonging to that type are shown underneath as thin lines. The mean radial velocity profile of all 
rays of that type, locally averaged in radius, is denoted by the series of numbers shown in each panel, in units of km/s 
(negative denoting inflow). The top inset above each panel plots the temperature derivative ${\rm d}T/{\rm d}r$, positive 
denoting increasing temperature with decreasing radius, for the prototypical ray (thick) and five other random sightlines 
of that type (thin).

\begin{table}
  \caption{The percentage of radial rays of each of the four types: `distant shock', `closer shock', `gradual heating', 
  and `low entropy'. The mean fraction across all eight haloes is calculated separately for each resolution level. Since 
  rays cover equal angular area, each fraction corresponds to the geometrical percentage of the sphere occupied by 
  sightlines satisfying each criterion. The errors represent the standard deviation among the eight haloes.} 
  \label{tRayFracs}
  \begin{center}
    \begin{tabular}{cccc}
     \hline 
       type & L9 & L10 & L11 \\ \hline\hline
       distant shock   & 23.6 $\pm$ 4.8\hphantom{0} & 17.3 $\pm$ 4.9\hphantom{0}  & 12.5 $\pm$ 6.1\hphantom{0}   \\
       closer shock    & 22.5 $\pm$ 15.9            & 27.2 $\pm$ 14.0             & 30.4 $\pm$ 15.2  \\
       gradual heating & 28.0 $\pm$ 15.2            & 14.5 $\pm$ 7.9\hphantom{0}  & 6.2  $\pm$ 3.9   \\
       low entropy     & 7.1  $\pm$ 2.7             & 7.5  $\pm$ 3.8              & 5.4  $\pm$ 2.9   \\ \hline
    \end{tabular}
  \end{center}
\end{table}

We briefly describe the behaviour of each type. Sightlines experiencing a `distant shock' undergo a jump in temperature 
from $\simeq$\,$10^4$\,K to $\simeq$\,$10^6$\,K over a radial range of $\sim$\,0.1$r_{\rm vir}$ (although, as in the 
example, multiple jumps can exist and be spread over a larger radial range). At this same radius the entropy also increases by 
approximately two orders of magnitude, from $\simeq$\,$10^{7.5}$\,K\,cm$^2$ to $\simeq$\,$10^{9.5}$\,K\,cm$^2$. The rapid 
inflow velocity decreases, reaching its maximum (positive) value near the radius of maximum temperature. Towards smaller 
radii, the ray temperature then follows the steadily increasing mean temperature profile of the halo, until reaching the 
centre. The `closer shock' sightlines have the same behaviour -- the radial distinction between close and distant shocks 
is arbitrary. In either case, the temperature profile can be non-monotonic if the shock occurs at sufficiently large 
radius, such that the shock temperature is greater than the local mean $T_{\rm gas}(r)$, and the gas can subsequently cool. 
Heating can be associated with one or multiple shocks (e.g. two, in the case of the prototypical example shown in the upper 
left panel). The maximum derivatives of temperature and entropy are $\simeq$\,2.0 (in log) per 0.1\,$r_{\rm vir}$, although 
the typical maximum along any given ray is roughly half as large.

The `gradual heating' sightlines generally reach the same maximum temperature (in the inner halo) and entropy (near the 
virialization boundary), but do so without any sudden jumps. Here the inflow velocity is roughly constant and always 
negative, never approaching a quasi-static state, increasing towards smaller radii 
and with a mean of approximately -75\,km/s. The `low entropy' rays are selected to have a maximum entropy of less than 
$8 \times 10^8$\,K\,cm$^2$, although the exact value is arbitrary. In this case, the inflow velocity is always 
strongly negative, peaking in the inner halo, with a mean of approximately -135\,km/s. The vast majority of these rays 
still reach high temperature, but at $\la$\,0.25$r_{\rm vir}$ and possibly not until the disk-halo interface. At 
$r > 0.5 r_{\rm vir}$ we see that `low entropy' sightlines also have systematically lower temperatures and 
higher inflow velocities than the other three types.

\begin{figure}
\centerline{\includegraphics[angle=0,width=3.4in]{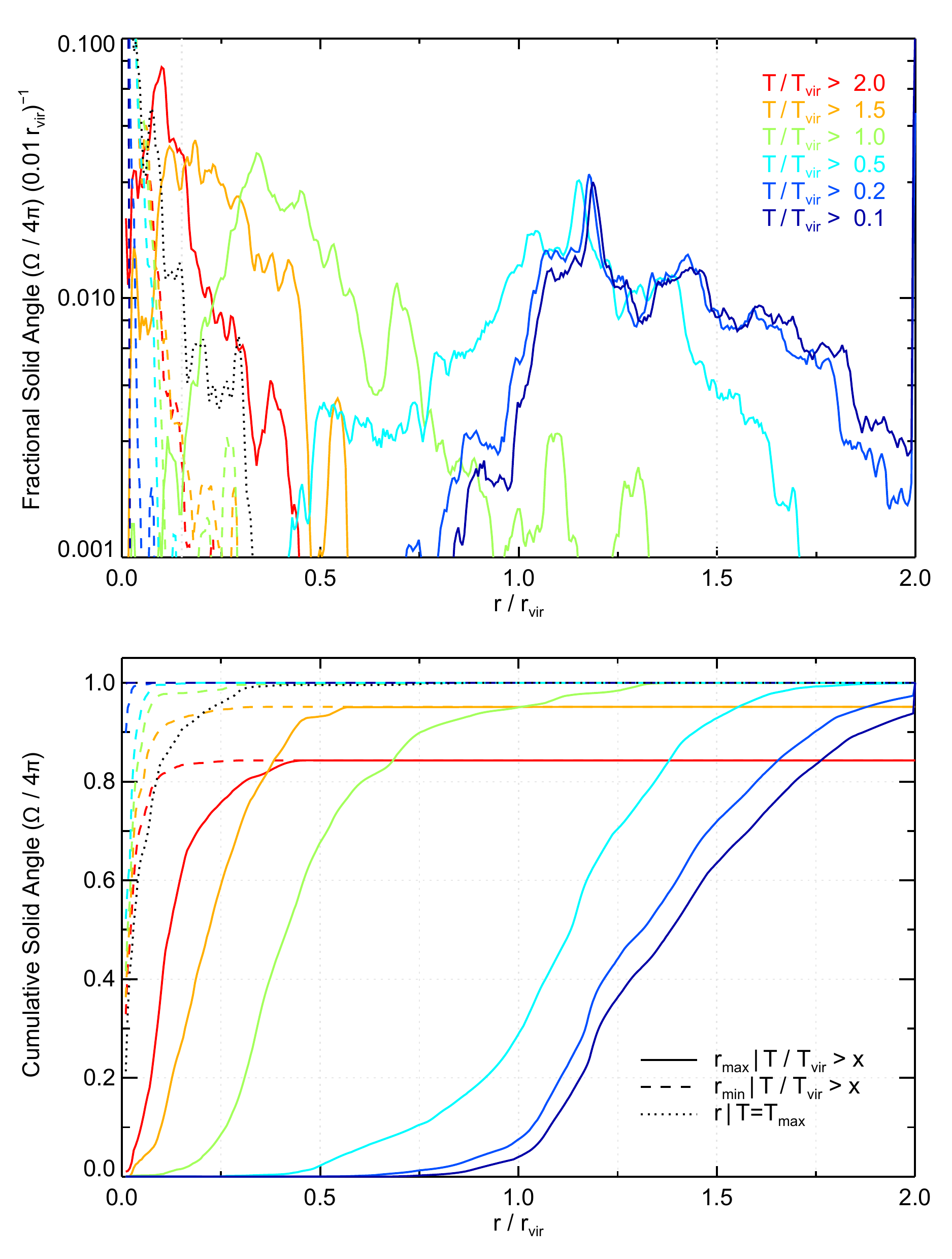}}
\caption{ (Top) The histogram of radii satisfying three different types of criteria (different linestyles), computed separately 
for each radial ray. Since each subtends equal angle, this histogram is equal to the fractional solid angle of the whole 
sphere covered by rays satisfying each criteria, as a (differential) function of radius. The criteria are: (i) the 
maximum radius at which the temperature along the ray exceeds some fraction of $T_{\rm vir}$, (ii) likewise, but the 
minimum radius, and (iii) the radius at which each ray reaches its maximum temperature. For the first two conditions, 
we consider six different temperature thresholds (from blue to red). In this figure we include only one halo at 
the high resolution level, h0L11, with $N_{\rm side}=64$ and $N_{\rm rad}=400$. (Bottom) As in the upper panel, 
but here as a cumulative function of radius.
 \label{fig_rayhisto_temp}} 
\end{figure}

In Table \ref{tRayFracs} we include the fraction of each of these four ray types, calculated as the mean over all eight 
haloes. Although the balance between the two shock types shifts with resolution, the total ray fraction experiencing a 
strong shock remains at $\simeq$\,45\%, with the temperature jumps moving somewhat inwards. The gradual heating fraction 
drops sharply at higher resolution levels, either because fluctuations become better resolved, or because the adopted cut 
on the temperature derivative is slightly too restrictive. The low entropy ray fraction remains fairly converged at 
$\simeq$\,6\%.

We have verified that these values, and all the other quantitative results related to the radial rays, are well converged 
with the numerical parameters $N_{\rm rad}=400$, $N_{\rm side}=64$. Given our selections, approximately $\simeq$\,15\% of 
the sightlines remain excluded from all four types at L9, although this increases to $\simeq$\,45\% at L11. The vast 
majority of these excluded rays either (i) fall just outside a selection, or (ii) experience a significant temperature 
dip within the halo. By definition they do not satisfy the `gradual heating' or `low entropy' conditions. 
Two such examples are shown as thin black lines in the upper two panels. We find that the narrow dips are typically a 
signature of intersecting satellite debris or satellite outskirts, which were not calculated as locally gravitationally 
bound and so were not excised. Gas cells of this type fill a larger fraction of the halo volume at higher resolution, 
leading to the increased percentage of rays excluded from all four types. Broader dips are typically intersections with 
large filamentary inflows which are not aligned in the radial direction. This exposes the main caveat of the 
above analysis -- namely, that gas inflow with a tangential velocity component need not evolve according to the 
temperature, entropy, or velocity structure of any particular radial ray. 

\subsection{Quantifying The Asphericity of Temperature}

In order to quantify the instantaneous halo temperature structure, we calculate differential histograms of radii 
satisfying a given criterion in Figure \ref{fig_rayhisto_temp} (top panel). First, solid lines show the 
distribution of the maximum radius, for each radial ray, where the temperature exceeds some fraction of $T_{\rm vir}$ 
(different colours). To avoid washing out features by stacking, we include only one halo (h0L11). We see that for the 
three lowest temperature cuts, $(0.1 - 0.5) \times T_{\rm vir}$, the $r_{\rm max}$ values all show a strong peak at 
$\simeq$\,1.2\,$r_{\rm vir}$, indicative of the virialization boundary. The distributions are broad, however. In the 
case of $T/T_{\rm vir} > 0.5$ the spread of maximum radii is roughly symmetric, extending from $0.5 r_{\rm vir}$ out to 
$1.6 r_{\rm vir}$. That is, although the majority of sightlines first exceed $0.5 T_{\rm vir}$ at a well-defined 
surface sitting at $1.2 r_{\rm vir}$, a non-negligible fraction first exceed this temperature already by 
$1.6 r_{\rm vir}$, and an even larger fraction cross this threshold inside the virial radius, between $0.5-0.75 r_{\rm vir}$.

The three higher temperature thresholds (green, orange, red) reflect the increasing mean radial temperature profile of 
the halo. The virial temperature is typically reached at $\simeq$\,0.5\,$r_{\rm vir}$, while gas reaches twice 
$T_{\rm vir}$ just prior to the galaxy at $\simeq$\,0.1\,$r_{\rm vir}$. The dashed lines show the minimum radius where 
each temperature threshold is reached. They are uniformly peaked at the halo centre. Finally, the dotted line plots the 
distribution of the radius where each ray reaches its maximum temperature, which are also peaked in the halo centre. 
One of our principal goals is to identify gas shocking at the halo-IGM transition. We see that the peak of the radial 
distribution, for rays satisfying $r_{\rm max}( T/T_{\rm vir} > \{0.5,0.2\} )$, measures the most dominant radius for 
this transition. With respect to halo to halo variability, we find that the shapes and overall widths of each $r_{\rm max}$ 
distribution can vary substantially. For instance, h0 has a strong peak which is both the narrowest in radius and the most 
covering in angular fraction. Yet, it also has a prominent secondary peak of roughly half the geometrical importance at 
larger radius, $\Delta r \simeq 0.2 r/r_{\rm vir}$ further from the halo centre. Even more extreme, h3 has two peaks of 
equal strength spaced $\Delta r \simeq 0.5 r/r_{\rm vir}$ apart. Multiple radial peaks are a common occurrence.

The widths of the distributions also encode a measure of the spherical symmetry of any strong heating. Here it is 
useful to look at the cumulative histograms of radii satisfying these same temperature criteria, as shown in the bottom 
panel of Figure \ref{fig_rayhisto_temp}. The slope of this CDF indicates angular uniformity -- presence of a Heaviside 
step function would indicate perfect spherical symmetry, while slopes approaching zero would indicate that the 
temperature transition takes place, depending on sightline, over widely disparate radii.
The fact that several temperature thresholds plateau below unity indicate that only $\simeq$\,85\% of rays 
reach $T > 2.0 T_{\rm vir}$ (red line) and $\simeq$\,95\% of rays reach $T > 1.5 T_{\rm vir}$ (orange line), while all 
other temperature thresholds are exceeded along all sightlines. Any value for the cumulative solid angle can provide a 
good way to quantify the location of the mean virialization boundary. For instance, the radius by which 2$\pi$ sr are 
covered by rays with maximum temperature exceeding half of $T_{\rm vir}$ is $\simeq$\,1.1$r_{\rm vir}$.

\begin{figure*}
\centerline{\includegraphics[angle=0,width=7.0in]{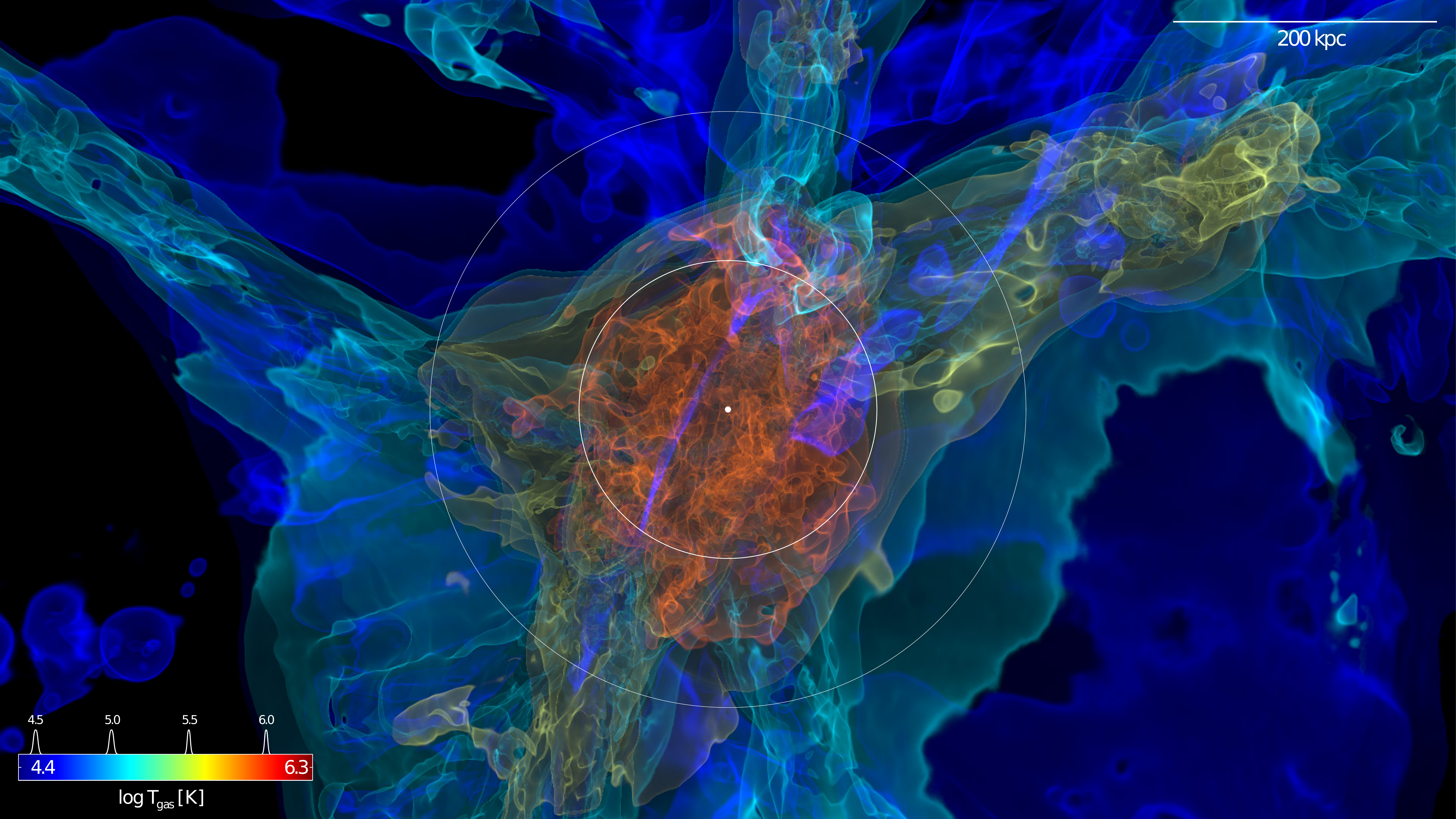}}
\caption{ A volume rendering of the three-dimensional temperature structure around a single halo (h0L11) at $z\!=\!2$. 
We use an orthographic projection with depth and height equal to 5.5\,$r_{\rm vir}$ ($\simeq$\,620\,kpc), as in the 
zoomed out panels of Figure \ref{fig_maps1b}. The colortable and 4-gaussian transfer function are shown. A front-to-back 
ray tracing method is used, with no scattering and no absorption. Along each ray we sample with a constant step-size of 
$0.25$ kpc. The temperature at each sample point is calculated using the standard cubic-spline SPH kernel interpolant with 
adaptive smoothing length $h$ over the $N=200$ nearest neighbours (the representation of the gas in terms of a Voronoi 
tessellation is not used). The white dot marks the halo centre, and the white circles denote one and two times the 
virial radius.
 \label{fig_volrend}} 
\end{figure*}

We conclude with a visual impression of the three-dimensional temperature structure in and around a single halo (h0L11) 
in Figure \ref{fig_volrend}. A ray-traced volume rendering highlights five iso-temperature surfaces using the same 
blue-red colortable and physical bounds of $4.4 < \log T_{\rm gas} \,[\rm{K}] < 6.3$ as in the temperature projections 
of Figure \ref{fig_maps1a}. It is sampled with a transfer function comprised of four narrow alpha-channel gaussians at 
$\log \,T_{\rm gas} \,[\rm{K}] = \{4.5,5.0,5.5,6.0\}$ with widths of $\sigma=\{6,6,3,3\} \times 10^{-3}$ in log K. 
The scale of the image is 5.5\,$r_{\rm vir}$ in both height and depth, as in the zoomed out panels of Figure 
\ref{fig_maps1b}. The dark blue surface straddles the mean $T_{\rm IGM}$ at $\simeq 2 r_{\rm vir}$ and reveals a 
morphology at overdensities just below cosmic web filaments. The light blue surface at $10^5$\,K clearly outlines the 
structure of Mpc-scale, large-scale filaments, while the yellow surface at $10^{5.5}$\,K is restricted to the interior 
of warmer filaments. Both of these two iso-temperatures surfaces are closed. Further, they enclose the virialised halo 
itself -- we can see how the filamentary shape of the $10^5$\,K surface at $r > 2 r_{\rm vir}$ transitions smoothly into 
a quasi-spherical shape by opening up, wrapping around and encompassing the higher temperature shells. The mean radius 
of the $10^{5.5}$\,K surface is $\simeq$\,1.0\,$r_{\rm vir}$, while that of the $10^6$\,K surface is 
$\simeq$\,0.5\,$r_{\rm vir}$. The intersection of radial sightlines cast outwards in different directions with this 
complex temperature structure in and around the $10^{12}$\,\msun halo gives rise to the four distinct ray types 
described earlier.


\section{Discussion} \label{sDiscussion}

\subsection{Observational Points}

We have deferred a direct comparison against observations of the gas content of haloes to future work. Such a comparison 
deserves a careful treatment of the various steps required to make robust synthetic (or `mock') observations from the 
simulations, for both hydrogen \citep[e.g.][]{fgk11,fg14,fum14,bird14} as well as metal signatures 
\citep[e.g.][]{shen12,hummels13,ford14,suresh15}. As motivation and in preparation for the next work in this series, 
there are a number of insightful observations of the gas in and around haloes with masses of 
$\sim$10$^{12}$\,\msun at $z\!\sim\!2$ which prompt mention here.

Observations have studied the incidence and kinematics of {\small HI} absorption around haloes of this mass 
\citep[e.g. the Lyman-break galaxies of][]{rudie12,rudie13}. They found, for instance, a covering fraction for Lyman-limit 
systems within the virial radius of $\sim$\,30\%. The hydrogen and metal content around lower mass damped 
Lyman-alpha systems (DLAs, $\sim$\,$10^{11}$\,\msun haloes) at this redshift is roughly consistent with their 
$10^{12}$\,\msun counterparts \citep{rubin14}, where the covering fractions within 100\,kpc of strong {\small Si II} 
absorption was found to be $\sim$\,20\%, versus $\sim$\,60\% for strong {\small C IV}.

Around more massive `quasar hosts' of 
$\sim$\,$10^{12.5}$\,\msun the covering fractions of neutral hydrogen are larger \citep{prochaska13}. On the scale of the 
virial radius in such haloes the covering fractions of cold metal ions \citep[e.g. {\small C II} and {\small C IV} 
in][]{prochaska14b} are even higher and may approach unity. The presence of cold gas at large radii \citep{turner14a} is 
a result which, in general, current hydrodynamical simulations have significant difficulty reproducing. On the other hand, 
hot gas as often probed by {\small O VI} at these redshifts \citep{turner14b} is more readily reproduced given sufficiently 
energetic outflows \citep{ford13,suresh15}. In some cases there is strong evidence that the observed absorption results 
from high velocity galactic outflows \citep{crighton15}. Other observations imply that slower inflowing structures with 
a high degree of kinematic coherence are more likely \citep{rubin14}. In general, it is unclear if the cold gas seen in 
the haloes surrounding galaxies is \textit{always} the consequence of one of (i) feedback-driven outflows 
\citep[e.g.][]{marinacci14b} or (ii) cosmological inflow, or if this depends on the halo mass under investigation. Either 
way, it is currently difficult, from the theoretical perspective, to explain how this cold gas is generated, distributed, 
and either maintained or replenished in the halo.

\subsection{The Resolution Issue}

Observations have also placed constraints on the physical size of the systems which give rise to metal absorption in 
gaseous haloes. \cite{simcoe06} estimated absorbers sizes around $z\!\sim\!2.3$ galaxies ranging from the sub-parsec to 
the kiloparsec. This is consistent with the transverse scales of {\small C IV} absorbers derived in \cite{rauch01} from 
multi-sightline analysis of lensed QSOs at sub-kpc scales \citep[see also][based on {\small Mg II}]{petitjean00}. Based 
on photo-ionization modelling of {\small C IV} absorbers at similar redshifts, \cite{schaye07} derived typical densities 
of $n_{\rm H} \sim 10^{-3.5}$\,cm$^{-3}$ with sizes of $\sim$\,100\,pc. Similarly, \cite{crighton15} observe metal 
absorption lines arising near a $z\!\simeq\!2.5$ galaxy and derive a $(<100-500)$\,pc size constraint, which they argue 
is a ubiquitous size-scale for all low-ion halo gas in the circumgalactic environments of both low and high redshift 
galaxies. \cite{pieri14} conclude that a typical low-ionization absorber associated with the CGM of strong Lyman-alpha 
forest systems at $z \sim 2.5$ is $\sim$\,30\,pc in size. Furthermore, the gas giving rise to various absorption lines 
may also have unexpected geometries with little resemblance to spherical clouds \citep{church14}.

Obtaining $\sim$\,100\,pc hydrodynamic resolution at the virial radius by simply running higher resolution simulations 
of isolated haloes will be prohibitively expensive. Following the observed scaling of the gas resolution in the halo, we 
would obtain a mean $r_{\rm cell}$\,$\simeq$\,0.1\,kpc physical at $z\!=\!2$ with a L14-class simulation. That is, gas 
mass resolution approaching 30\,\msun, requiring $\ga$\,$10^{10}$ gas cells in a $10^{12}$\,\msun halo. This is not a 
realistic goal for the near future. Even given such a simulation, the minimum size of resolved clouds would be a few 
times larger than the size of individual resolution elements.

On the other hand, numerical methods may enable novel ways to focus resolution within the halo regime. If we can avoid 
clustering the computational effort in the densest regions of space -- within galaxies themselves -- much higher spatial 
resolutions can be achieved in lower density media. In AMR simulations, the refinement criterion can be chosen as 
desired, on the gradient of the neutral hydrogen fraction for instance \citep{rosdahl12}, or with a progressively refined 
uniform Eulerian grid which is volume filling \citep{miniati14}. In our moving mesh simulations, 
we can enforce a target gas cell mass criterion of any type through adaptive refinement and de-refinement of Voronoi 
cells. In future work we will explore this possibility as a solution to the resolution issue, in addition to quantifying 
the rates and modes of accretion using the tracer particle information.


\section{Conclusions} \label{sConclusions}

In this first paper of the series we present a suite of high-resolution cosmological hydrodynamic zoom-in 
simulations targeted at understanding the properties of halo gas and cosmological accretion in $\sim$\,$10^{12}$\,\msun 
haloes at $z\!=\!2$. Using the moving-mesh code {\small AREPO} we simulate each of eight haloes at three levels of 
increasing resolution, reaching a mean baryonic mass of $\sim$10,000 solar masses. We study the thermal and dynamical 
state of halo gas within 2\,$r_{\rm vir}$ and in particular quantify density, temperature, entropy, angular momentum, 
and radial velocity in terms of their mean radial profiles and, more interestingly, in terms of their variation along 
radial sightlines in different directions.

In this mass regime, haloes typically reside at the intersection of one or more `cosmic web' filaments 
arising from large-scale structure. The result is a significant amount of filamentary inflow across the virial radius. 

\begin{itemize}
\item
Investigating the interaction of this inflow and the quasi-static hot halo atmosphere, we find that stream morphologies 
become continuously more complex with better numerical resolution. In general, single coherent flows tend to resolve 
into multiple, narrower streams while producing density and temperature structure at smaller spatial scales. However, we 
point out that even at our highest resolution -- comparable to the best zoom-in simulations that currently exist for 
haloes of this mass -- the gas-dynamics in the circumgalactic regime are poorly resolved in comparison to within galaxies 
themselves. 

\item
In general, the mean radial profiles of gas properties are well captured at our lowest resolution -- equivalent to that 
currently available in full cosmological volume simulations. However, gas which is stripped from infalling satellites, 
or which forms as the result of a strong tidal interaction, is less well converged, and populates the halo with 
additional high density gas in our highest resolution runs.

\item
Although average radial profiles are well converged, their direct interpretation can be misleading. Within the halo 
itself, $0.2 < r/r_{\rm vir} < 1.0$, there are clearly multiple gas components overlapping in radius and corresponding 
to quasi-static versus inflowing material. For example, although the mean radial velocity within the halo is near zero 
(equilibrium), this is not true for the majority of gas, which is actually either inflowing or outflowing. 

\item
Examining the thermal and dynamical state of gas within 2\,$r_{\rm vir}$, we clearly identify the existence of a 
strong virial shock. This `virialization boundary' typically resides at $(1.25 - 1.5) \times r_{\rm vir}$ and is evident 
as a sharp increase in gas temperature, entropy, density, and a decrease in inwards radial velocity. Collimated inflows 
which remain cold past the virial radius experience significant heating, typically at $\sim$\,0.5$r_{\rm vir}$, from 
$\la$\,10$^{4.5}$\,K to $\ga$\,10$^6$\,K.

\item
Although the mean radius of a strong virial shock may be $\simeq$\,1.25\,$r_{\rm vir}$, our sightline analysis allows 
us to identify, in each halo, many radially-sharp virial shocks, each at some radius between 1.0\,$r_{\rm vir}$ and 
1.5\,$r_{\rm vir}$, depending on angle on the virial sphere.

\item
Investigating the process of gas virialization, we identify different mechanisms responsible for the heating of gas in 
the circumgalactic regime. In addition to a single strong virial shock at $\la$\,$r_{\rm vir}$, we find that gas can shock 
at much larger radii, $\la$\,2\,$r_{\rm vir}$, particularly in systems where the hot halo is far from equilibrium and can 
extend to these distances due to the dynamical response of a recent major merger. We also see that gas can heat gradually, 
with temperature increasing slowly from its characteristic IGM value until reaching the mean halo temperature profile. 
Finally, we identify the existence of radial sightlines along which gas entropy remains always below the level 
characteristic of the hot halo. Even in this case, however, gas cannot avoid heating to $\sim$\,$T_{\rm vir}$ in the inner 
halo, $\sim$\,0.2\,$r_{\rm vir}$, just prior to cooling onto the centrally forming galaxy. Both of these two cases -- 
gradual heating and persistent low entropy -- are sub-dominant, accounting geometrically for only $\sim$\,15\% of the 
total 4$\pi$, with a strong shock covering the other $\sim$\,85\%.

\item
Finally, we conclude by assessing the ashpericity of halo gas and by measuring the location and width of the virialization 
boundary as demarcated by temperature. We find that the distributions of maximum radius $r_{\rm max}$, where the 
temperature exceeds e.g. $(0.1 - 0.5) \times T_{\rm vir}$ along each radial sightline, can be quite broad. In general, 
for the less disturbed systems, the majority of radial sightlines first surpasses $0.5$\,$T_{\rm vir}$ at a well-defined 
surface sitting at or just beyond $r_{\rm vir}$. However, a non-negligible number already exceeds this temperature at larger 
distance, and an even larger fraction crosses this threshold only inside the virial radius, between $0.5-0.75 r_{\rm vir}$. 
We propose that this geometrically-motivated analysis can be used to quantify the structure of gas heating as a result 
of the halo-IGM transition.
\end{itemize}

In order to serve as a benchmark for the realization of these eight haloes, we have intentionally not included 
existing models for energetic feedback due to star formation (in the form of a kinetic galactic-scale wind generation) 
and AGN (in the form of quasar and radio mode thermal energy input plus local radiative effects). We have deferred any 
direct comparisons with observations of the gas content of haloes. Both points remain directions for future work, 
particularly in the sense of how simulations \textit{without} galactic winds can reproduce observed hydrogen and metal 
signatures. The answer may depend sensitively on details of the numerics, including hydrodynamic resolution, 
particularly if the spatial scale of dense gas structures in galactic haloes is as small as commonly estimated.


\section*{Acknowledgements.}

DN would like to thank Vicente Rodriguez-Gomez for allowing us to use the {\small SUBLINK} 
merger tree code prior to its release, and Joshua Suresh for many useful discussions, comments, and suggestions. 
The computations presented in this paper were performed on the Odyssey cluster at Harvard University.
SG acknowledges support for program number HST-HF2-51341.001-A provided by NASA through a Hubble Fellowship 
grant from the STScI, which is operated by the Association of Universities for Research in Astronomy, Incorporated, 
under NASA contract NAS5-26555. 
VS acknowledges support by the European Research Council under ERC-StG grant EXAGAL-308037. 
LH acknowledges support from NASA grant NNX12AC67G and NSF grant AST-1312095.

\bibliographystyle{mn2e}
\bibliography{refs}

\begin{thebibliography}{}
\makeatletter
\relax
\def\mn@urlcharsother{\let\do\@makeother \do\$\do\&\do\#\do\^\do\_\do\%\do\~}
\def\mn@doi{\begingroup\mn@urlcharsother \@ifnextchar[{\mn@doi@}{\mn@doi@[]}}
\def\mn@doi@[#1]#2{\def\@tempa{#1}\ifx\@tempa\@empty
  \href{http://dx.doi.org/#2}{doi:#2}\else \href{http://dx.doi.org/#2}{#1}\fi
  \endgroup}
\def\mn@eprint#1#2{\mn@eprint@#1:#2::\@nil}
\def\mn@eprint@arXiv#1{\href{http://arxiv.org/abs/#1}{{\tt arXiv:#1}}}
\def\mn@eprint@dblp#1{\href{http://dblp.uni-trier.de/rec/bibtex/#1.xml}{dblp:#1}}
\def\mn@eprint@#1:#2:#3:#4\@nil{\def\@tempa {#1}\def\@tempb {#2}\def\@tempc
  {#3}\ifx \@tempc \@empty \let\@tempc\@tempb \let\@tempb\@tempa \fi \ifx
  \@tempb \@empty \def\@tempb{arXiv}\fi \@ifundefined
  {mn@eprint@\@tempb}{\@tempb:\@tempc}{\expandafter \expandafter \csname
  mn@eprint@\@tempb\endcsname \expandafter{\@tempc}}}

\bibitem[\protect\citeauthoryear{{Abadi}, {Navarro}, {Steinmetz}  \&
  {Eke}}{{Abadi} et~al.}{2003}]{abadi03}
{Abadi} M.~G.,  {Navarro} J.~F.,  {Steinmetz} M.,   {Eke} V.~R.,  2003, \mn@doi
  [\apj] {10.1086/375512}, \href
  {http://adsabs.harvard.edu/abs/2003ApJ...591..499A} {591, 499}

\bibitem[\protect\citeauthoryear{{Agertz}, {Teyssier}  \& {Moore}}{{Agertz}
  et~al.}{2009}]{agertz09}
{Agertz} O.,  {Teyssier} R.,   {Moore} B.,  2009, \mn@doi [\mnras]
  {10.1111/j.1745-3933.2009.00685.x}, \href
  {http://adsabs.harvard.edu/abs/2009MNRAS.397L..64A} {397, L64}

\bibitem[\protect\citeauthoryear{{Aragon-Calvo}, {Neyrinck}  \&
  {Silk}}{{Aragon-Calvo} et~al.}{2014}]{aragon15}
{Aragon-Calvo} M.~A.,  {Neyrinck} M.~C.,   {Silk} J.,  2014, ArXiv e-prints,
  \href {http://adsabs.harvard.edu/abs/2014arXiv1412.1119A} {}

\bibitem[\protect\citeauthoryear{{Barkana} \& {Loeb}}{{Barkana} \&
  {Loeb}}{2001}]{bl01}
{Barkana} R.,  {Loeb} A.,  2001, \mn@doi [\physrep]
  {10.1016/S0370-1573(01)00019-9}, \href
  {http://adsabs.harvard.edu/abs/2001PhR...349..125B} {349, 125}

\bibitem[\protect\citeauthoryear{{Barnes} \& {Efstathiou}}{{Barnes} \&
  {Efstathiou}}{1987}]{barnes87}
{Barnes} J.,  {Efstathiou} G.,  1987, \mn@doi [\apj] {10.1086/165480}, \href
  {http://adsabs.harvard.edu/abs/1987ApJ...319..575B} {319, 575}

\bibitem[\protect\citeauthoryear{{Barnes} \& {Hut}}{{Barnes} \&
  {Hut}}{1986}]{bh86}
{Barnes} J.,  {Hut} P.,  1986, \mn@doi [\nat] {10.1038/324446a0}, \href
  {http://adsabs.harvard.edu/abs/1986Natur.324..446B} {324, 446}

\bibitem[\protect\citeauthoryear{{Bird}, {Vogelsberger}, {Haehnelt}, {Sijacki},
  {Genel}, {Torrey}, {Springel}  \& {Hernquist}}{{Bird} et~al.}{2014}]{bird14}
{Bird} S.,  {Vogelsberger} M.,  {Haehnelt} M.,  {Sijacki} D.,  {Genel} S.,
  {Torrey} P.,  {Springel} V.,   {Hernquist} L.,  2014, \mn@doi [\mnras]
  {10.1093/mnras/stu1923}, \href
  {http://adsabs.harvard.edu/abs/2014MNRAS.445.2313B} {445, 2313}

\bibitem[\protect\citeauthoryear{{Birnboim} \& {Dekel}}{{Birnboim} \&
  {Dekel}}{2003a}]{bd03}
{Birnboim} Y.,  {Dekel} A.,  2003a, \mn@doi [\mnras]
  {10.1046/j.1365-8711.2003.06955.x}, \href
  {http://adsabs.harvard.edu/abs/2003MNRAS.345..349B} {345, 349}

\bibitem[\protect\citeauthoryear{{Birnboim} \& {Dekel}}{{Birnboim} \&
  {Dekel}}{2003b}]{birnboim03}
{Birnboim} Y.,  {Dekel} A.,  2003b, \mn@doi [\mnras]
  {10.1046/j.1365-8711.2003.06955.x}, \href
  {http://adsabs.harvard.edu/abs/2003MNRAS.345..349B} {345, 349}

\bibitem[\protect\citeauthoryear{{Birnboim}, {Dekel}  \& {Neistein}}{{Birnboim}
  et~al.}{2007}]{birnboim07}
{Birnboim} Y.,  {Dekel} A.,   {Neistein} E.,  2007, \mn@doi [\mnras]
  {10.1111/j.1365-2966.2007.12074.x}, \href
  {http://adsabs.harvard.edu/abs/2007MNRAS.380..339B} {380, 339}

\bibitem[\protect\citeauthoryear{{Bullock}, {Dekel}, {Kolatt}, {Kravtsov},
  {Klypin}, {Porciani}  \& {Primack}}{{Bullock} et~al.}{2001}]{bullock01}
{Bullock} J.~S.,  {Dekel} A.,  {Kolatt} T.~S.,  {Kravtsov} A.~V.,  {Klypin}
  A.~A.,  {Porciani} C.,   {Primack} J.~R.,  2001, \mn@doi [\apj]
  {10.1086/321477}, \href {http://adsabs.harvard.edu/abs/2001ApJ...555..240B}
  {555, 240}

\bibitem[\protect\citeauthoryear{{Churchill}, {Vander Vliet}, {Trujillo-Gomez},
  {Kacprzak}  \& {Klypin}}{{Churchill} et~al.}{2014}]{church14}
{Churchill} C.~W.,  {Vander Vliet} J.~R.,  {Trujillo-Gomez} S.,  {Kacprzak}
  G.~G.,   {Klypin} A.,  2014, ArXiv e-prints, \href
  {http://adsabs.harvard.edu/abs/2014arXiv1409.0914C} {}

\bibitem[\protect\citeauthoryear{{Crighton}, {Hennawi}, {Simcoe}, {Cooksey},
  {Murphy}, {Fumagalli}, {Prochaska}  \& {Shanks}}{{Crighton}
  et~al.}{2015}]{crighton15}
{Crighton} N.~H.~M.,  {Hennawi} J.~F.,  {Simcoe} R.~A.,  {Cooksey} K.~L.,
  {Murphy} M.~T.,  {Fumagalli} M.,  {Prochaska} J.~X.,   {Shanks} T.,  2015,
  \mn@doi [\mnras] {10.1093/mnras/stu2088}, \href
  {http://adsabs.harvard.edu/abs/2015MNRAS.446...18C} {446, 18}

\bibitem[\protect\citeauthoryear{{Danovich}, {Dekel}, {Hahn}  \&
  {Teyssier}}{{Danovich} et~al.}{2012}]{danovich12}
{Danovich} M.,  {Dekel} A.,  {Hahn} O.,   {Teyssier} R.,  2012, \mn@doi
  [\mnras] {10.1111/j.1365-2966.2012.20751.x}, \href
  {http://adsabs.harvard.edu/abs/2012MNRAS.422.1732D} {422, 1732}

\bibitem[\protect\citeauthoryear{{Danovich}, {Dekel}, {Hahn}, {Ceverino}  \&
  {Primack}}{{Danovich} et~al.}{2014}]{danovich14}
{Danovich} M.,  {Dekel} A.,  {Hahn} O.,  {Ceverino} D.,   {Primack} J.,  2014,
  preprint, (arXiv:1407.7129), \href
  {http://adsabs.harvard.edu/abs/2014arXiv1407.7129D} {}

\bibitem[\protect\citeauthoryear{{Dekel} et~al.,}{{Dekel}
  et~al.}{2009}]{dekel09}
{Dekel} A.,  et~al., 2009, \mn@doi [\nat] {10.1038/nature07648}, \href
  {http://adsabs.harvard.edu/abs/2009Natur.457..451D} {457, 451}

\bibitem[\protect\citeauthoryear{{Dolag}, {Borgani}, {Murante}  \&
  {Springel}}{{Dolag} et~al.}{2009}]{dolag09}
{Dolag} K.,  {Borgani} S.,  {Murante} G.,   {Springel} V.,  2009, \mn@doi
  [\mnras] {10.1111/j.1365-2966.2009.15034.x}, \href
  {http://adsabs.harvard.edu/abs/2009MNRAS.399..497D} {399, 497}

\bibitem[\protect\citeauthoryear{{Dubois} et~al.,}{{Dubois}
  et~al.}{2014}]{dubois14}
{Dubois} Y.,  et~al., 2014, \mn@doi [\mnras] {10.1093/mnras/stu1227}, \href
  {http://adsabs.harvard.edu/abs/2014MNRAS.444.1453D} {444, 1453}

\bibitem[\protect\citeauthoryear{{Faucher-Gigu{\`e}re}, {Lidz}, {Zaldarriaga}
  \& {Hernquist}}{{Faucher-Gigu{\`e}re} et~al.}{2009}]{fg09}
{Faucher-Gigu{\`e}re} C.-A.,  {Lidz} A.,  {Zaldarriaga} M.,   {Hernquist} L.,
  2009, \mn@doi [\apj] {10.1088/0004-637X/703/2/1416}, \href
  {http://adsabs.harvard.edu/abs/2009ApJ...703.1416F} {703, 1416}

\bibitem[\protect\citeauthoryear{{Faucher-Gigu{\`e}re}, {Kere{\v s}}  \&
  {Ma}}{{Faucher-Gigu{\`e}re} et~al.}{2011}]{fgk11}
{Faucher-Gigu{\`e}re} C.-A.,  {Kere{\v s}} D.,   {Ma} C.-P.,  2011, \mn@doi
  [\mnras] {10.1111/j.1365-2966.2011.19457.x}, \href
  {http://adsabs.harvard.edu/abs/2011MNRAS.417.2982F} {417, 2982}

\bibitem[\protect\citeauthoryear{{Faucher-Giguere}, {Hopkins}, {Keres},
  {Muratov}, {Quataert}  \& {Murray}}{{Faucher-Giguere} et~al.}{2014}]{fg14}
{Faucher-Giguere} C.-A.,  {Hopkins} P.~F.,  {Keres} D.,  {Muratov} A.~L.,
  {Quataert} E.,   {Murray} N.,  2014, ArXiv e-prints, \href
  {http://adsabs.harvard.edu/abs/2014arXiv1409.1919F} {}

\bibitem[\protect\citeauthoryear{{Feldmann} \& {Mayer}}{{Feldmann} \&
  {Mayer}}{2015}]{feldmann15}
{Feldmann} R.,  {Mayer} L.,  2015, \mn@doi [\mnras] {10.1093/mnras/stu2207},
  \href {http://adsabs.harvard.edu/abs/2015MNRAS.446.1939F} {446, 1939}

\bibitem[\protect\citeauthoryear{{Ford}, {Oppenheimer}, {Dav{\'e}}, {Katz},
  {Kollmeier}  \& {Weinberg}}{{Ford} et~al.}{2013}]{ford13}
{Ford} A.~B.,  {Oppenheimer} B.~D.,  {Dav{\'e}} R.,  {Katz} N.,  {Kollmeier}
  J.~A.,   {Weinberg} D.~H.,  2013, \mn@doi [\mnras] {10.1093/mnras/stt393},
  \href {http://adsabs.harvard.edu/abs/2013MNRAS.432...89F} {432, 89}

\bibitem[\protect\citeauthoryear{{Ford}, {Dav{\'e}}, {Oppenheimer}, {Katz},
  {Kollmeier}, {Thompson}  \& {Weinberg}}{{Ford} et~al.}{2014}]{ford14}
{Ford} A.~B.,  {Dav{\'e}} R.,  {Oppenheimer} B.~D.,  {Katz} N.,  {Kollmeier}
  J.~A.,  {Thompson} R.,   {Weinberg} D.~H.,  2014, \mn@doi [\mnras]
  {10.1093/mnras/stu1418}, \href
  {http://adsabs.harvard.edu/abs/2014MNRAS.444.1260F} {444, 1260}

\bibitem[\protect\citeauthoryear{{Fumagalli}, {Hennawi}, {Prochaska}, {Kasen},
  {Dekel}, {Ceverino}  \& {Primack}}{{Fumagalli} et~al.}{2014}]{fum14}
{Fumagalli} M.,  {Hennawi} J.~F.,  {Prochaska} J.~X.,  {Kasen} D.,  {Dekel} A.,
   {Ceverino} D.,   {Primack} J.,  2014, \mn@doi [\apj]
  {10.1088/0004-637X/780/1/74}, \href
  {http://adsabs.harvard.edu/abs/2014ApJ...780...74F} {780, 74}

\bibitem[\protect\citeauthoryear{{Gabor} \& {Bournaud}}{{Gabor} \&
  {Bournaud}}{2014}]{gabor14}
{Gabor} J.~M.,  {Bournaud} F.,  2014, \mn@doi [\mnras] {10.1093/mnrasl/slt139},
  \href {http://adsabs.harvard.edu/abs/2014MNRAS.437L..56G} {437, L56}

\bibitem[\protect\citeauthoryear{{Gabor} \& {Dav{\'e}}}{{Gabor} \&
  {Dav{\'e}}}{2012}]{gabor12}
{Gabor} J.~M.,  {Dav{\'e}} R.,  2012, \mn@doi [\mnras]
  {10.1111/j.1365-2966.2012.21640.x}, \href
  {http://adsabs.harvard.edu/abs/2012MNRAS.427.1816G} {427, 1816}

\bibitem[\protect\citeauthoryear{{Genel}, {Vogelsberger}, {Nelson}, {Sijacki},
  {Springel}  \& {Hernquist}}{{Genel} et~al.}{2013}]{genel13}
{Genel} S.,  {Vogelsberger} M.,  {Nelson} D.,  {Sijacki} D.,  {Springel} V.,
  {Hernquist} L.,  2013, \mn@doi [\mnras] {10.1093/mnras/stt1383}, \href
  {http://adsabs.harvard.edu/abs/2013MNRAS.435.1426G} {435, 1426}

\bibitem[\protect\citeauthoryear{{George}, {Fabian}, {Sanders}, {Young}  \&
  {Russell}}{{George} et~al.}{2009}]{george09}
{George} M.~R.,  {Fabian} A.~C.,  {Sanders} J.~S.,  {Young} A.~J.,   {Russell}
  H.~R.,  2009, \mn@doi [\mnras] {10.1111/j.1365-2966.2009.14547.x}, \href
  {http://adsabs.harvard.edu/abs/2009MNRAS.395..657G} {395, 657}

\bibitem[\protect\citeauthoryear{{G{\'o}rski}, {Hivon}, {Banday}, {Wandelt},
  {Hansen}, {Reinecke}  \& {Bartelmann}}{{G{\'o}rski} et~al.}{2005}]{gorski05}
{G{\'o}rski} K.~M.,  {Hivon} E.,  {Banday} A.~J.,  {Wandelt} B.~D.,  {Hansen}
  F.~K.,  {Reinecke} M.,   {Bartelmann} M.,  2005, \mn@doi [\apj]
  {10.1086/427976}, \href {http://adsabs.harvard.edu/abs/2005ApJ...622..759G}
  {622, 759}

\bibitem[\protect\citeauthoryear{{Hahn} \& {Abel}}{{Hahn} \&
  {Abel}}{2011}]{hahn11}
{Hahn} O.,  {Abel} T.,  2011, \mn@doi [\mnras]
  {10.1111/j.1365-2966.2011.18820.x}, \href
  {http://adsabs.harvard.edu/abs/2011MNRAS.415.2101H} {415, 2101}

\bibitem[\protect\citeauthoryear{{Hahn} \& {Angulo}}{{Hahn} \&
  {Angulo}}{2015}]{hahn15}
{Hahn} O.,  {Angulo} R.,  2015, preprint, (arXiv:1501.01959), \href
  {http://adsabs.harvard.edu/abs/2015arXiv150101959H} {}

\bibitem[\protect\citeauthoryear{{Hennawi} et~al.,}{{Hennawi}
  et~al.}{2006}]{hennawi06}
{Hennawi} J.~F.,  et~al., 2006, \mn@doi [\apj] {10.1086/507069}, \href
  {http://adsabs.harvard.edu/abs/2006ApJ...651...61H} {651, 61}

\bibitem[\protect\citeauthoryear{{Hummels}, {Bryan}, {Smith}  \&
  {Turk}}{{Hummels} et~al.}{2013}]{hummels13}
{Hummels} C.~B.,  {Bryan} G.~L.,  {Smith} B.~D.,   {Turk} M.~J.,  2013, \mn@doi
  [\mnras] {10.1093/mnras/sts702}, \href
  {http://adsabs.harvard.edu/abs/2013MNRAS.430.1548H} {430, 1548}

\bibitem[\protect\citeauthoryear{{Iapichino}, {Adamek}, {Schmidt}  \&
  {Niemeyer}}{{Iapichino} et~al.}{2008}]{iapichino08}
{Iapichino} L.,  {Adamek} J.,  {Schmidt} W.,   {Niemeyer} J.~C.,  2008, \mn@doi
  [\mnras] {10.1111/j.1365-2966.2008.13137.x}, \href
  {http://adsabs.harvard.edu/abs/2008MNRAS.388.1079I} {388, 1079}

\bibitem[\protect\citeauthoryear{{Joung}, {Putman}, {Bryan}, {Fern{\'a}ndez}
  \& {Peek}}{{Joung} et~al.}{2012}]{joung12}
{Joung} M.~R.,  {Putman} M.~E.,  {Bryan} G.~L.,  {Fern{\'a}ndez} X.,   {Peek}
  J.~E.~G.,  2012, \mn@doi [\apj] {10.1088/0004-637X/759/2/137}, \href
  {http://adsabs.harvard.edu/abs/2012ApJ...759..137J} {759, 137}

\bibitem[\protect\citeauthoryear{{Katz}, {Weinberg}  \& {Hernquist}}{{Katz}
  et~al.}{1996}]{katz96}
{Katz} N.,  {Weinberg} D.~H.,   {Hernquist} L.,  1996, \mn@doi [\apjs]
  {10.1086/192305}, \href {http://adsabs.harvard.edu/abs/1996ApJS..105...19K}
  {105, 19}

\bibitem[\protect\citeauthoryear{{Katz}, {Keres}, {Dave}  \& {Weinberg}}{{Katz}
  et~al.}{2003}]{katz03}
{Katz} N.,  {Keres} D.,  {Dave} R.,   {Weinberg} D.~H.,  2003, in {Rosenberg}
  J.~L.,  {Putman} M.~E.,  eds,  Astrophysics and Space Science Library Vol.
  281, The IGM/Galaxy Connection. The Distribution of Baryons at z=0. p.~185,
  \mn@eprint {} {arXiv:astro-ph/0209279}

\bibitem[\protect\citeauthoryear{{Kere{\v s}}, {Katz}, {Weinberg}  \&
  {Dav{\'e}}}{{Kere{\v s}} et~al.}{2005}]{keres05}
{Kere{\v s}} D.,  {Katz} N.,  {Weinberg} D.~H.,   {Dav{\'e}} R.,  2005, \mn@doi
  [\mnras] {10.1111/j.1365-2966.2005.09451.x}, \href
  {http://adsabs.harvard.edu/abs/2005MNRAS.363....2K} {363, 2}

\bibitem[\protect\citeauthoryear{{Kere{\v s}}, {Katz}, {Fardal}, {Dav{\'e}}  \&
  {Weinberg}}{{Kere{\v s}} et~al.}{2009}]{keres09}
{Kere{\v s}} D.,  {Katz} N.,  {Fardal} M.,  {Dav{\'e}} R.,   {Weinberg} D.~H.,
  2009, \mn@doi [\mnras] {10.1111/j.1365-2966.2009.14541.x}, \href
  {http://adsabs.harvard.edu/abs/2009MNRAS.395..160K} {395, 160}

\bibitem[\protect\citeauthoryear{{Khandai}, {Di Matteo}, {Croft}, {Wilkins},
  {Feng}, {Tucker}, {DeGraf}  \& {Liu}}{{Khandai} et~al.}{2014}]{khandai14}
{Khandai} N.,  {Di Matteo} T.,  {Croft} R.,  {Wilkins} S.~M.,  {Feng} Y.,
  {Tucker} E.,  {DeGraf} C.,   {Liu} M.-S.,  2014, preprint, (arXiv:1402.0888),
  \href {http://adsabs.harvard.edu/abs/2014arXiv1402.0888K} {}

\bibitem[\protect\citeauthoryear{{Lewis}, {Challinor}  \& {Lasenby}}{{Lewis}
  et~al.}{2000}]{lewis00}
{Lewis} A.,  {Challinor} A.,   {Lasenby} A.,  2000, \mn@doi [\apj]
  {10.1086/309179}, \href {http://adsabs.harvard.edu/abs/2000ApJ...538..473L}
  {538, 473}

\bibitem[\protect\citeauthoryear{{Makino}, {Sasaki}  \& {Suto}}{{Makino}
  et~al.}{1998}]{makino98}
{Makino} N.,  {Sasaki} S.,   {Suto} Y.,  1998, \mn@doi [\apj] {10.1086/305507},
  \href {http://adsabs.harvard.edu/abs/1998ApJ...497..555M} {497, 555}

\bibitem[\protect\citeauthoryear{{Marinacci}, {Pakmor}  \&
  {Springel}}{{Marinacci} et~al.}{2014a}]{marinacci14a}
{Marinacci} F.,  {Pakmor} R.,   {Springel} V.,  2014a, \mn@doi [\mnras]
  {10.1093/mnras/stt2003}, \href
  {http://adsabs.harvard.edu/abs/2014MNRAS.437.1750M} {437, 1750}

\bibitem[\protect\citeauthoryear{{Marinacci}, {Pakmor}, {Springel}  \&
  {Simpson}}{{Marinacci} et~al.}{2014b}]{marinacci14b}
{Marinacci} F.,  {Pakmor} R.,  {Springel} V.,   {Simpson} C.~M.,  2014b,
  \mn@doi [\mnras] {10.1093/mnras/stu1136}, \href
  {http://adsabs.harvard.edu/abs/2014MNRAS.442.3745M} {442, 3745}

\bibitem[\protect\citeauthoryear{{Miniati}}{{Miniati}}{2014}]{miniati14}
{Miniati} F.,  2014, \mn@doi [\apj] {10.1088/0004-637X/782/1/21}, \href
  {http://adsabs.harvard.edu/abs/2014ApJ...782...21M} {782, 21}

\bibitem[\protect\citeauthoryear{{Muratov}, {Keres}, {Faucher-Giguere},
  {Hopkins}, {Quataert}  \& {Murray}}{{Muratov} et~al.}{2015}]{muratov15}
{Muratov} A.~L.,  {Keres} D.,  {Faucher-Giguere} C.-A.,  {Hopkins} P.~F.,
  {Quataert} E.,   {Murray} N.,  2015, ArXiv e-prints, \href
  {http://adsabs.harvard.edu/abs/2015arXiv150103155M} {}

\bibitem[\protect\citeauthoryear{{Nelson}, {Vogelsberger}, {Genel}, {Sijacki},
  {Kere{\v s}}, {Springel}  \& {Hernquist}}{{Nelson} et~al.}{2013}]{nelson13}
{Nelson} D.,  {Vogelsberger} M.,  {Genel} S.,  {Sijacki} D.,  {Kere{\v s}} D.,
  {Springel} V.,   {Hernquist} L.,  2013, \mn@doi [\mnras]
  {10.1093/mnras/sts595}, \href
  {http://adsabs.harvard.edu/abs/2013MNRAS.429.3353N} {429, 3353}

\bibitem[\protect\citeauthoryear{{Nelson}, {Genel}, {Vogelsberger}, {Springel},
  {Sijacki}, {Torrey}  \& {Hernquist}}{{Nelson} et~al.}{2015}]{nelson15a}
{Nelson} D.,  {Genel} S.,  {Vogelsberger} M.,  {Springel} V.,  {Sijacki} D.,
  {Torrey} P.,   {Hernquist} L.,  2015, \mn@doi [\mnras]
  {10.1093/mnras/stv017}, \href
  {http://adsabs.harvard.edu/abs/2015MNRAS.448...59N} {448, 59}

\bibitem[\protect\citeauthoryear{{O{\~n}orbe}, {Garrison-Kimmel}, {Maller},
  {Bullock}, {Rocha}  \& {Hahn}}{{O{\~n}orbe} et~al.}{2014}]{onorbe14}
{O{\~n}orbe} J.,  {Garrison-Kimmel} S.,  {Maller} A.~H.,  {Bullock} J.~S.,
  {Rocha} M.,   {Hahn} O.,  2014, \mn@doi [\mnras] {10.1093/mnras/stt2020},
  \href {http://adsabs.harvard.edu/abs/2014MNRAS.437.1894O} {437, 1894}

\bibitem[\protect\citeauthoryear{{Ocvirk}, {Pichon}  \& {Teyssier}}{{Ocvirk}
  et~al.}{2008}]{ocvirk08}
{Ocvirk} P.,  {Pichon} C.,   {Teyssier} R.,  2008, \mn@doi [\mnras]
  {10.1111/j.1365-2966.2008.13763.x}, \href
  {http://adsabs.harvard.edu/abs/2008MNRAS.390.1326O} {390, 1326}

\bibitem[\protect\citeauthoryear{{Oppenheimer}, {Dav{\'e}}, {Kere{\v s}},
  {Fardal}, {Katz}, {Kollmeier}  \& {Weinberg}}{{Oppenheimer}
  et~al.}{2010}]{opp10}
{Oppenheimer} B.~D.,  {Dav{\'e}} R.,  {Kere{\v s}} D.,  {Fardal} M.,  {Katz}
  N.,  {Kollmeier} J.~A.,   {Weinberg} D.~H.,  2010, \mn@doi [\mnras]
  {10.1111/j.1365-2966.2010.16872.x}, \href
  {http://adsabs.harvard.edu/abs/2010MNRAS.406.2325O} {406, 2325}

\bibitem[\protect\citeauthoryear{{Petitjean}, {Aracil}, {Srianand}  \&
  {Ibata}}{{Petitjean} et~al.}{2000}]{petitjean00}
{Petitjean} P.,  {Aracil} B.,  {Srianand} R.,   {Ibata} R.,  2000, \aap, \href
  {http://adsabs.harvard.edu/abs/2000A%26A...359..457P} {359, 457}

\bibitem[\protect\citeauthoryear{{Pieri} et~al.,}{{Pieri}
  et~al.}{2014}]{pieri14}
{Pieri} M.~M.,  et~al., 2014, \mn@doi [\mnras] {10.1093/mnras/stu577}, \href
  {http://adsabs.harvard.edu/abs/2014MNRAS.441.1718P} {441, 1718}

\bibitem[\protect\citeauthoryear{{Prochaska}, {Hennawi}  \&
  {Simcoe}}{{Prochaska} et~al.}{2013}]{prochaska13}
{Prochaska} J.~X.,  {Hennawi} J.~F.,   {Simcoe} R.~A.,  2013, \mn@doi [\apjl]
  {10.1088/2041-8205/762/2/L19}, \href
  {http://adsabs.harvard.edu/abs/2013ApJ...762L..19P} {762, L19}

\bibitem[\protect\citeauthoryear{{Prochaska}, {Lau}  \& {Hennawi}}{{Prochaska}
  et~al.}{2014}]{prochaska14b}
{Prochaska} J.~X.,  {Lau} M.~W.,   {Hennawi} J.~F.,  2014, \mn@doi [\apj]
  {10.1088/0004-637X/796/2/140}, \href
  {http://adsabs.harvard.edu/abs/2014ApJ...796..140P} {796, 140}

\bibitem[\protect\citeauthoryear{{Putman}, {Peek}  \& {Joung}}{{Putman}
  et~al.}{2012}]{putman12}
{Putman} M.~E.,  {Peek} J.~E.~G.,   {Joung} M.~R.,  2012, \mn@doi [\araa]
  {10.1146/annurev-astro-081811-125612}, \href
  {http://adsabs.harvard.edu/abs/2012ARA%26A..50..491P} {50, 491}

\bibitem[\protect\citeauthoryear{{Rauch}, {Sargent}, {Barlow}  \&
  {Carswell}}{{Rauch} et~al.}{2001}]{rauch01}
{Rauch} M.,  {Sargent} W.~L.~W.,  {Barlow} T.~A.,   {Carswell} R.~F.,  2001,
  \mn@doi [\apj] {10.1086/323523}, \href
  {http://adsabs.harvard.edu/abs/2001ApJ...562...76R} {562, 76}

\bibitem[\protect\citeauthoryear{{Rees} \& {Ostriker}}{{Rees} \&
  {Ostriker}}{1977}]{rees77}
{Rees} M.~J.,  {Ostriker} J.~P.,  1977, \mnras, \href
  {http://adsabs.harvard.edu/abs/1977MNRAS.179..541R} {179, 541}

\bibitem[\protect\citeauthoryear{{Rodriguez-Gomez} et~al.,}{{Rodriguez-Gomez}
  et~al.}{2015}]{rodrig15}
{Rodriguez-Gomez} V.,  et~al., 2015, ArXiv e-prints, \href
  {http://adsabs.harvard.edu/abs/2015arXiv150201339R} {}

\bibitem[\protect\citeauthoryear{{Rosdahl} \& {Blaizot}}{{Rosdahl} \&
  {Blaizot}}{2012}]{rosdahl12}
{Rosdahl} J.,  {Blaizot} J.,  2012, \mn@doi [\mnras]
  {10.1111/j.1365-2966.2012.20883.x}, \href
  {http://adsabs.harvard.edu/abs/2012MNRAS.423..344R} {423, 344}

\bibitem[\protect\citeauthoryear{{Rubin}, {Hennawi}, {Prochaska}, {Simcoe},
  {Myers}  \& {Wingyee Lau}}{{Rubin} et~al.}{2014}]{rubin14}
{Rubin} K.~H.~R.,  {Hennawi} J.~F.,  {Prochaska} J.~X.,  {Simcoe} R.~A.,
  {Myers} A.,   {Wingyee Lau} M.,  2014, ArXiv e-prints, \href
  {http://adsabs.harvard.edu/abs/2014arXiv1411.6016R} {}

\bibitem[\protect\citeauthoryear{{Rudie} et~al.,}{{Rudie}
  et~al.}{2012}]{rudie12}
{Rudie} G.~C.,  et~al., 2012, \mn@doi [\apj] {10.1088/0004-637X/750/1/67},
  \href {http://adsabs.harvard.edu/abs/2012ApJ...750...67R} {750, 67}

\bibitem[\protect\citeauthoryear{{Rudie}, {Steidel}, {Shapley}  \&
  {Pettini}}{{Rudie} et~al.}{2013}]{rudie13}
{Rudie} G.~C.,  {Steidel} C.~C.,  {Shapley} A.~E.,   {Pettini} M.,  2013,
  \mn@doi [\apj] {10.1088/0004-637X/769/2/146}, \href
  {http://adsabs.harvard.edu/abs/2013ApJ...769..146R} {769, 146}

\bibitem[\protect\citeauthoryear{{S{\'a}nchez Almeida}, {Elmegreen},
  {Mu{\~n}oz-Tu{\~n}{\'o}n}  \& {Elmegreen}}{{S{\'a}nchez Almeida}
  et~al.}{2014}]{almeida14}
{S{\'a}nchez Almeida} J.,  {Elmegreen} B.~G.,  {Mu{\~n}oz-Tu{\~n}{\'o}n} C.,
  {Elmegreen} D.~M.,  2014, \mn@doi [\aapr] {10.1007/s00159-014-0071-1}, \href
  {http://adsabs.harvard.edu/abs/2014A%26ARv..22...71S} {22, 71}

\bibitem[\protect\citeauthoryear{{Scannapieco} et~al.,}{{Scannapieco}
  et~al.}{2012}]{scan12a}
{Scannapieco} C.,  et~al., 2012, \mn@doi [\mnras]
  {10.1111/j.1365-2966.2012.20993.x}, \href
  {http://adsabs.harvard.edu/abs/2012MNRAS.423.1726S} {423, 1726}

\bibitem[\protect\citeauthoryear{{Schaal} \& {Springel}}{{Schaal} \&
  {Springel}}{2015}]{schaal15}
{Schaal} K.,  {Springel} V.,  2015, \mn@doi [\mnras] {10.1093/mnras/stu2386},
  \href {http://adsabs.harvard.edu/abs/2015MNRAS.446.3992S} {446, 3992}

\bibitem[\protect\citeauthoryear{{Schaye}, {Carswell}  \& {Kim}}{{Schaye}
  et~al.}{2007}]{schaye07}
{Schaye} J.,  {Carswell} R.~F.,   {Kim} T.-S.,  2007, \mn@doi [\mnras]
  {10.1111/j.1365-2966.2007.12005.x}, \href
  {http://adsabs.harvard.edu/abs/2007MNRAS.379.1169S} {379, 1169}

\bibitem[\protect\citeauthoryear{{Schaye} et~al.,}{{Schaye}
  et~al.}{2015}]{schaye15}
{Schaye} J.,  et~al., 2015, \mn@doi [\mnras] {10.1093/mnras/stu2058}, \href
  {http://adsabs.harvard.edu/abs/2015MNRAS.446..521S} {446, 521}

\bibitem[\protect\citeauthoryear{{Shen}, {Madau}, {Guedes}, {Mayer},
  {Prochaska}  \& {Wadsley}}{{Shen} et~al.}{2013}]{shen12}
{Shen} S.,  {Madau} P.,  {Guedes} J.,  {Mayer} L.,  {Prochaska} J.~X.,
  {Wadsley} J.,  2013, \mn@doi [\apj] {10.1088/0004-637X/765/2/89}, \href
  {http://adsabs.harvard.edu/abs/2013ApJ...765...89S} {765, 89}

\bibitem[\protect\citeauthoryear{{Silk}}{{Silk}}{1977}]{silk77}
{Silk} J.,  1977, \mn@doi [\apj] {10.1086/154972}, \href
  {http://adsabs.harvard.edu/abs/1977ApJ...211..638S} {211, 638}

\bibitem[\protect\citeauthoryear{{Simcoe}, {Sargent}, {Rauch}  \&
  {Becker}}{{Simcoe} et~al.}{2006}]{simcoe06}
{Simcoe} R.~A.,  {Sargent} W.~L.~W.,  {Rauch} M.,   {Becker} G.,  2006, \mn@doi
  [\apj] {10.1086/498441}, \href
  {http://adsabs.harvard.edu/abs/2006ApJ...637..648S} {637, 648}

\bibitem[\protect\citeauthoryear{{Springel}}{{Springel}}{2005}]{spr05b}
{Springel} V.,  2005, \mn@doi [\mnras] {10.1111/j.1365-2966.2005.09655.x},
  \href {http://adsabs.harvard.edu/abs/2005MNRAS.364.1105S} {364, 1105}

\bibitem[\protect\citeauthoryear{{Springel}}{{Springel}}{2010}]{spr10}
{Springel} V.,  2010, \mn@doi [\mnras] {10.1111/j.1365-2966.2009.15715.x}, 401,
  791

\bibitem[\protect\citeauthoryear{{Springel} \& {Hernquist}}{{Springel} \&
  {Hernquist}}{2003}]{spr03}
{Springel} V.,  {Hernquist} L.,  2003, \mn@doi [\mnras]
  {10.1046/j.1365-8711.2003.06206.x}, 339, 289

\bibitem[\protect\citeauthoryear{{Springel}, {White}, {Tormen}  \&
  {Kauffmann}}{{Springel} et~al.}{2001}]{spr01}
{Springel} V.,  {White} S.~D.~M.,  {Tormen} G.,   {Kauffmann} G.,  2001,
  \mn@doi [\mnras] {10.1046/j.1365-8711.2001.04912.x}, \href
  {http://adsabs.harvard.edu/abs/2001MNRAS.328..726S} {328, 726}

\bibitem[\protect\citeauthoryear{{Steidel}, {Erb}, {Shapley}, {Pettini},
  {Reddy}, {Bogosavljevi{\'c}}, {Rudie}  \& {Rakic}}{{Steidel}
  et~al.}{2010}]{steidel10}
{Steidel} C.~C.,  {Erb} D.~K.,  {Shapley} A.~E.,  {Pettini} M.,  {Reddy} N.,
  {Bogosavljevi{\'c}} M.,  {Rudie} G.~C.,   {Rakic} O.,  2010, \mn@doi [\apj]
  {10.1088/0004-637X/717/1/289}, \href
  {http://adsabs.harvard.edu/abs/2010ApJ...717..289S} {717, 289}

\bibitem[\protect\citeauthoryear{{Suresh}, {Bird}, {Vogelsberger}, {Genel},
  {Torrey}, {Sijacki}, {Springel}  \& {Hernquist}}{{Suresh}
  et~al.}{2015}]{suresh15}
{Suresh} J.,  {Bird} S.,  {Vogelsberger} M.,  {Genel} S.,  {Torrey} P.,
  {Sijacki} D.,  {Springel} V.,   {Hernquist} L.,  2015, preprint,
  (arXiv:1501.02267), \href
  {http://adsabs.harvard.edu/abs/2015arXiv1501.02267A} {}

\bibitem[\protect\citeauthoryear{{Suto}, {Sasaki}  \& {Makino}}{{Suto}
  et~al.}{1998}]{suto98}
{Suto} Y.,  {Sasaki} S.,   {Makino} N.,  1998, \mn@doi [\apj] {10.1086/306520},
  509, 544

\bibitem[\protect\citeauthoryear{{Turner}, {Schaye}, {Steidel}, {Rudie}  \&
  {Strom}}{{Turner} et~al.}{2014a}]{turner14b}
{Turner} M.~L.,  {Schaye} J.,  {Steidel} C.~C.,  {Rudie} G.~C.,   {Strom}
  A.~L.,  2014a, ArXiv e-prints, \href
  {http://adsabs.harvard.edu/abs/2014arXiv1410.8214T} {}

\bibitem[\protect\citeauthoryear{{Turner}, {Schaye}, {Steidel}, {Rudie}  \&
  {Strom}}{{Turner} et~al.}{2014b}]{turner14a}
{Turner} M.~L.,  {Schaye} J.,  {Steidel} C.~C.,  {Rudie} G.~C.,   {Strom}
  A.~L.,  2014b, \mn@doi [\mnras] {10.1093/mnras/stu1801}, \href
  {http://adsabs.harvard.edu/abs/2014MNRAS.445..794T} {445, 794}

\bibitem[\protect\citeauthoryear{{Vogelsberger}, {Sijacki}, {Kere{\v s}},
  {Springel}  \& {Hernquist}}{{Vogelsberger} et~al.}{2012}]{vog12}
{Vogelsberger} M.,  {Sijacki} D.,  {Kere{\v s}} D.,  {Springel} V.,
  {Hernquist} L.,  2012, \mn@doi [\mnras] {10.1111/j.1365-2966.2012.21590.x},
  \href {http://adsabs.harvard.edu/abs/2012MNRAS.425.3024V} {425, 3024}

\bibitem[\protect\citeauthoryear{{Vogelsberger} et~al.,}{{Vogelsberger}
  et~al.}{2014}]{vog14a}
{Vogelsberger} M.,  et~al., 2014, \mn@doi [\nat] {10.1038/nature13316}, \href
  {http://adsabs.harvard.edu/abs/2014Natur.509..177V} {509, 177}

\bibitem[\protect\citeauthoryear{{Wetzel} \& {Nagai}}{{Wetzel} \&
  {Nagai}}{2014}]{wetzel15}
{Wetzel} A.~R.,  {Nagai} D.,  2014, ArXiv e-prints, \href
  {http://adsabs.harvard.edu/abs/2014arXiv1412.0662W} {}

\bibitem[\protect\citeauthoryear{{White} \& {Frenk}}{{White} \&
  {Frenk}}{1991}]{wf91}
{White} S.~D.~M.,  {Frenk} C.~S.,  1991, \mn@doi [\apj] {10.1086/170483}, \href
  {http://adsabs.harvard.edu/abs/1991ApJ...379...52W} {379, 52}

\bibitem[\protect\citeauthoryear{{White} \& {Rees}}{{White} \&
  {Rees}}{1978}]{wr78}
{White} S.~D.~M.,  {Rees} M.~J.,  1978, \mnras, \href
  {http://adsabs.harvard.edu/abs/1978MNRAS.183..341W} {183, 341}

\bibitem[\protect\citeauthoryear{{Yoshikawa}, {Yoshida}  \&
  {Umemura}}{{Yoshikawa} et~al.}{2013}]{yoshikawa13}
{Yoshikawa} K.,  {Yoshida} N.,   {Umemura} M.,  2013, \mn@doi [\apj]
  {10.1088/0004-637X/762/2/116}, \href
  {http://adsabs.harvard.edu/abs/2013ApJ...762..116Y} {762, 116}

\bibitem[\protect\citeauthoryear{{Zavala}, {Balogh}, {Afshordi}  \&
  {Ro}}{{Zavala} et~al.}{2012}]{zavala12}
{Zavala} J.,  {Balogh} M.~L.,  {Afshordi} N.,   {Ro} S.,  2012, \mn@doi
  [\mnras] {10.1111/j.1365-2966.2012.21980.x}, \href
  {http://adsabs.harvard.edu/abs/2012MNRAS.426.3464Z} {426, 3464}

\bibitem[\protect\citeauthoryear{{van Leer}}{{van Leer}}{1977}]{vl77}
{van Leer} B.,  1977, \mn@doi [J. Comput. Phys.]
  {10.1016/0021-9991(77)90095-X}, \href
  {http://adsabs.harvard.edu/abs/1977JCoPh..23..276V} {23, 276}

\bibitem[\protect\citeauthoryear{{van de Voort} \& {Schaye}}{{van de Voort} \&
  {Schaye}}{2012}]{vdv12b}
{van de Voort} F.,  {Schaye} J.,  2012, \mn@doi [\mnras]
  {10.1111/j.1365-2966.2012.20949.x}, \href
  {http://adsabs.harvard.edu/abs/2012MNRAS.tmp.2882V} {p.~2882}

\makeatother
\end{thebibliography}

\end{document}